 \documentclass[prc,amsmath,amsfonts,showpacs,eqsecnum,floatfix,twocolumn]{revtex4}
 \usepackage{bm}         
 \usepackage{graphicx}  
 \usepackage{pstcol}    
 \usepackage{axodraw}   
 
\begin{document}
\hyphenation{Rijken}
\hyphenation{Nijmegen}
 
\title{
    Extended-soft-core baryon-baryon model \\   
    III. $S=-2$ hyperon-hyperon/nucleon interaction }
\author{Th.A.\ Rijken}
\email[]{t.rijken@science.ru.nl}
\affiliation{ Institute of Mathematics, Astrophysics, and Particle Physics \\
 University of Nijmegen, The Netherlands}

\author{Y.\ Yamamoto}
\email[]{yamamoto@tsuru.ac.jp}  
\affiliation{Physics Section, Tsuru University, Tsuru, Yamanashi 402-8555, Japan}


\date{version of: \today}
 
\begin{abstract}
This paper presents the Extended-Soft-Core (ESC) potentials ESC04a-ESC04d 
for baryon-baryon channels with total strangeness $S=-2$.          
For these channels no experimental scattering data exist, and 
also the information from hypernuclei is very limited.
The potential models for $S=-2$ are
based on SU(3) extensions of potential models for the $S=0$ and $S=-1$
sectors, which {\it are\/} fitted to experimental data. 
Flavor SU(3)-symmetry is broken 'kinematically' by the masses   
of the baryons and the mesons. Moreover, in ESC04a,b also
the coupling constants are broken, albeit in a well defined way using the
$^3P_0$ quark-antiquark pair creation model as a guidance.
But, the fit to the $S=0$ and $S=-1$ sectors
provides the necessary constraints to fix all free parameters.
Therefore, the potentials for the $S=-2$ sectors 
do not contain additional free parameters, which situation is similar to
the soft-core  one-boson-exchange NSC97-models.              
Various properties of the potentials are illustrated by giving results
for scattering lengths, bound states, phase-parameters, and total cross sections.

The features of $\Xi$ hypernuclei predicted by ESC04d are studied on the basis
of the G-matrix approach.
\end{abstract}

\maketitle

\section{Introduction}
In this paper the Extended-Soft-Core (ESC) potentials ESC04a-ESC04d, described
in the companion nucleon-nucleon ({\it N\!N}) \cite{Rij05} and the hyperon-nucleon 
({\it Y\!N})
paper \cite{RY05}, for baryon-baryon channels with total strangeness $S=-2$.          
These papers will be referred to as paper I and II respectively.
In \cite{SR99} the Nijmegen soft-core one-boson-exchange (OBE) interactions
NSC97a-f for baryon-baryon (BB) systems for $S=-2,-3,-4$ were presented.

For these channels hardly any experimental scattering information is available, 
and also the information from hypernuclei is very limited. There are data on
double $\Lambda\Lambda$-hypernuclei, which recently became very much improved
by the observation of the Nagara-event \cite{Tak01}. This event indicates that the 
$\Lambda\Lambda$-interaction is rather weak, in contrast to the estimates
based on the older experimental observations \cite{Dan63,Pro66}.

In the virtual absense of experimental information, we assume that 
the potentials obey (broken) flavor SU(3) symmetry. As in I and II,
the potentials are parametrized in terms of meson-baryon-baryon, 
and meson-pair-baryon-baryon couplings and gaussian form factors.
This enables us to include in the interaction one-boson-exchange (OBE),
two-pseudoscalar-exchange (TME), and meson-pair-exchange (MPE), 
without any new parameters. All parameters have been fixed by a simultaneous 
fit to the {\it N\!N} and {\it Y\!N} data, described in I and II. 
Each ${\it N\!N} \oplus {\it YN}$-model
leads to a {\it Y\!Y}-model in a well defined way. In II we have introduced four
different models, called ESC04a-d, based on the options: SU(3)-symmetry
breaking/ no-breaking
of coupling constants, and pure pv/ pv-ps mixture for the pseudoscalar
meson couplings. 
Then, SU(3)-symmetry allows us to define all coupling constants
needed to describe the multi-strange interactions
in the baryon-baryon channels occurring in $\{8\}\otimes\{8\}$.
Most of the details on the SU(3) description are well known, and in particular
for baryon-baryon scattering they can be found in papers I, II, and e.g.
\cite{MRS89,RSY99,SR99}. So, here we restrict ourselves to a minimal
exposition of these matters, necessary for the readability of this paper.
Therefore, in Sec.~\ref{sec:2} 
we first review for $S=-2$ the baryon-baryon multi-channel description, and 
present the SU(3)-symmetric interaction Lagrangian describing
the interaction vertices between mesons and members of the
$J^P={\textstyle     (1/2) }^+$ baryon octet, and define their coupling
constants.  We then identify the various channels which 
occur in the $S=-2$ baryon-baryon systems. 
In appendix~\ref{app:C} the potentials on the isospin basis are given in terms
of the SU(3)-irreps.
In most cases, the
interaction is a multi-channel interaction, characterized by transition
potentials and thresholds. Details were given in \cite{SR99,RSY99}.            
For the details on the pair-interactions, we refer to II \cite{RY05}.
In Sec.~\ref{sec:exch} we give a general treatment of the problem
of flavor-exchange forces, which is very helpful to understand 
the proper treatment of exchange forces and the treatment 
of baryon-baryon channels with identical particles.
In Sec.~\ref{sec:thresh} we describe briefly the treatment of the 
multi-channel threshods in the potentials.
In Sec.~\ref{sec:results} we present the results of the ESC04
potentials for all the sectors with total strangeness $S=-2$.
We give the couplings and $F/(F+D)$-ratio's for OBE-exchanges of ESC04a,d.
Similarly, tables with the pair-couplings are shown in appendix~\ref{app:A}.
We give the $S$-wave scattering lengths, discuss the possibility of
bound states in these partial waves. Also, we give the S-matrix information for the
elastic channels in terms of the Bryan-Klarsfeld-Sprung (BKS) phase parameters 
\cite{Bry81,Kla83,Spr85}, or in the Kabir-Kermode (KK) \cite{Kab87} format.
Tables with the BKS-phase parameters are displayed in appendix~\ref{app:B}.                
Such information is very usefull for example for the
construction of the $\Lambda$-, $\Sigma$-, and $\Xi$-nucleus potentials.
We also give results for the total cross sections for all leading channels.

Important differences among the four versions ESC04a,b,c,d appear in their
$\Xi N$ sectors. Table XXV in Ref.~\cite{RY05} demonstrates that ESC04a,b 
(ESC04c,d) lead to repulsive (attractive) $\Xi$ potentials in nuclear matter.
Especially, the $\Xi N$ interaction of ESC04d is attractive enough to
produce various $\Xi$ hypernuclei. It is very interesting to study their 
features on the basis of the G-matrix approach. In Sec.~\ref{sec:gmat1}, we represent
the $\Xi N$ G-matrix interactions derived from ESC04d as density-dependent
local potentials.
In Sect.~\ref{sec:gmat2}, structure calculations for $\Xi$ hypernuclei are performed
with use of $\Xi$-nucleus folding potentials obtained from the G-matrix
interactions. It is discussed how the features of ESC04d appear in the 
level structure of $\Xi$ hypernuclei.
We conclude the paper with a summary and some final remarks in Sec.~\ref{sec:conc}.

\section{Channels, Potentials, and SU(3) Symmetry}
\label{sec:2}

\subsection{Multi-channel Formalism}     
\label{sec:2a}
In this paper we consider the baryon-baryon reactions with $S=-2$
\begin{equation}
 A_1(p_a,s_a)+B_1(p_b,s_b) \rightarrow A_2(p'_a,s'_a)+B_2(p'_b,s'_b)             
\label{eq:2.1}\end{equation}
Like in Ref.'s~\cite{MRS89,RSY99} we will for the {\it YN}-channels 
also refer to $A_1$ and $A_2$
as particles 1 and 3, and to $B_1$ and $B_2$ as particles 2 and 4. For the 
kinematics and the definition of the amplitudes, we refer to paper I 
\cite{Rij05} of this series. Similar material can be found in \cite{MRS89}.
Also, in paper I the derivation of the Lippmann-Schwinger equation 
in the context of the relativistic two-body equation is described.

On the physical particle basis, there are four charge channels:
\begin{eqnarray}
   q=+2:\ \  && \Sigma^+\Sigma^+\rightarrow\Sigma^+\Sigma^+,    \nonumber\\
   q=+1:\ \  && (\Xi^0 p, \Sigma^+\Lambda, \Sigma^0\Sigma^+)\rightarrow
             (\Xi^0 p, \Sigma^+\Lambda, \Sigma^0\Sigma^+),      \nonumber\\
   q=\ \ 0:\ \ && (\Lambda\Lambda , \Xi^0n, \Xi^-p, 
                   \Sigma^0\Lambda, \Sigma^0\Sigma^0, \Sigma^-\Sigma^+)\rightarrow
   \nonumber\\ && (\Lambda\Lambda , \Xi^0n, \Xi^-p, 
                   \Sigma^0\Lambda, \Sigma^0\Sigma^0, \Sigma^-\Sigma^+), \nonumber\\
   q=-1:\ \  && (\Xi^- n, \Sigma^-\Lambda, \Sigma^-\Sigma^0)\rightarrow
             (\Xi^- n, \Sigma^-\Lambda, \Sigma^-\Sigma^0),      \nonumber\\
   q=-2:\ \  && \Sigma^-\Sigma^-\rightarrow\Sigma^-\Sigma^-.
\label{eq:2.2}\end{eqnarray}
Like in \cite{MRS89,RSY99}, the potentials are calculated on the isospin basis.
For $S=-2$ hyperon-nucleon systems there are three isospin channels:
\begin{eqnarray}
I={\textstyle 0}:\ \ && (\Lambda \Lambda, \Xi N, \Sigma \Sigma\rightarrow
          \Lambda \Lambda, \xi N, \Sigma \Sigma), \nonumber\\      
I={\textstyle 1}:\ \ && (\Xi N, \Sigma\Lambda, \Sigma \Sigma\rightarrow
                           \Xi N, \Sigma\Lambda, \Sigma \Sigma), \nonumber\\
I={\textstyle 2}:\ \ && \Sigma \Sigma\rightarrow\Sigma\Sigma.      
\label{eq:2.3}\end{eqnarray}

  \begin{figure}   
  \resizebox{7.25cm}{8.75cm}
  {\includegraphics[100,500][500,850]{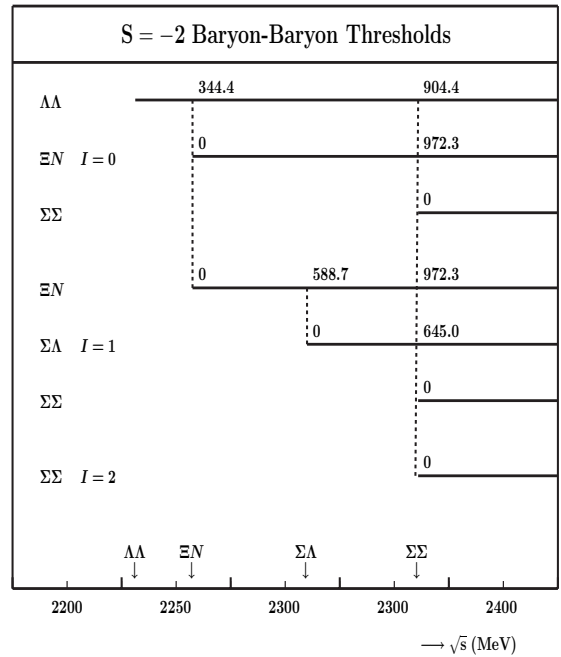}}
 \caption{Thresholds in {\it YN}- and {\it YY}-channels for $S=-2$. }                
 \label{thresh1}
  \end{figure}

For the kinematics of the reactions and the various thresholds, see \cite{RSY99}.
In this work we do not solve the Lippmann-Schwinger equation, but the 
multi-channel Schr\"{o}dinger equation in configuration space, completely 
analogous to \cite{MRS89}.
The multi-channel Schr\"odinger equation for the configuration-space
potential is derived from the Lippmann-Schwinger equation through
the standard Fourier transform, and the equation for the radial
wave function is found to be of the form~\cite{MRS89}
\begin{equation}
   u^{\prime\prime}_{l,j}+(p_i^2\delta_{i,j}-A_{i,j})u_{l,j}
       -B_{i,j}u'_{l,j}=0,           
\label{eq:2.6}\end{equation}
where $A_{i,j}$ contains the potential, nonlocal contributions, and
the centrifugal barrier, while $B_{i,j}$ is only present when non-local
contributions are included. 
The solution in the presence of open and closed channels is given,
for example, in Ref.~\cite{Nag73}.
The inclusion of the Coulomb interaction in the
configuration-space equation is well known and included in the evaluation
of the scattering matrix.

Obviously, the potential on the particle basis for the $q=2$ and
$q=-2$ channels are given by the $I={\textstyle 2}$
$\Sigma\Sigma $ potential on the isospin basis. For $q=0$ and $q=\pm 1$, the
potentials are related to the potentials on the isospin basis by an
isospin rotation. 
Using the indices $a, b, c, d$ for $\Lambda\Lambda, \Xi N, \Lambda\Sigma$, and
$\Sigma\Sigma$ respectively, we have \cite{Mae95}
 \begin{widetext}
 \begin{equation}
 V(q=0) =
  \left(\begin{array}{cccccc}
    V_{aa}
    &  \sqrt{\textstyle\frac{1}{2}}V_{ba}             
    & -\sqrt{\textstyle\frac{1}{2}}V_{ba}             
    & 0
    & -\sqrt{\textstyle\frac{1}{3}}V_{ad}             
    &  \sqrt{\textstyle\frac{1}{3}}V_{ad} \\[2mm]     
      \cdot
    &  {\textstyle\frac{1}{2}}\left[V_{bb}(1)+V_{bb}(0)\right]                
    &  {\textstyle\frac{1}{2}}\left[V_{bb}(1)-V_{bb}(0)\right]                
    &  \sqrt{\textstyle\frac{1}{2}}V_{bc}             
    & -\sqrt{\textstyle\frac{1}{6}}V_{bd}(0)          
    &  \sqrt{\textstyle\frac{1}{6}}V_{bd}(0)                                 
      -{\textstyle\frac{1}{2}}V_{bd}(1) \\[2mm]                         
      \cdot & \cdot
    &  {\textstyle\frac{1}{2}}\left[V_{bb}(1)+V_{bb}(0)\right]                
    &  \sqrt{\textstyle\frac{1}{2}}V_{bc}             
    &  \sqrt{\textstyle\frac{1}{6}}V_{bd}(0)          
    & -\sqrt{\textstyle\frac{1}{6}}V_{bd}(0)                                 
      -{\textstyle\frac{1}{2}}V_{bd}(1) \\[2mm]                         
      \cdot & \cdot & \cdot & V_{cc} & 0  
    &  -\sqrt{\textstyle\frac{1}{2}}V_{cd} \\[2mm]     
      \cdot & \cdot & \cdot & \cdot         
    &  {\textstyle\frac{1}{3}}\left[2V_{dd}(2)+V_{dd}(0)\right]                
    &  {\textstyle\frac{1}{3}}\left[2V_{dd}(2)-V_{dd}(0)\right] \\[2mm]        
      \cdot & \cdot & \cdot & \cdot & \cdot 
    &  {\textstyle\frac{1}{6}}\left[V_{dd}(2)+3V_{dd}(1)+2V_{dd}(0)\right]                
        \end{array}\right)\ ,
\label{eq:2.4} \end{equation}
and for $q=+1$ we have
 \begin{equation}
 V(q=+1) =
  \left(\begin{array}{ccc}
    V_{bb}(1) & V_{bc} & -\sqrt{\textstyle\frac{1}{2}}V_{bd} \\[2mm]     
    V_{bc}    & V_{cc} & -\sqrt{\textstyle\frac{1}{2}}V_{cd} \\[2mm]     
      -\sqrt{\textstyle\frac{1}{2}}V_{bd}(1)          
    & -\sqrt{\textstyle\frac{1}{2}}V_{cd}          
    &  {\textstyle\frac{1}{2}}\left[V_{dd}(1)+V_{dd}(2)\right]                
        \end{array}\right)\ ,
\label{eq:2.5} \end{equation}
 \end{widetext}
Here, when necessary an isospin label is added in parentheses.

The momentum space and configuration space potentials for the ESC04-model
have been described in paper I \cite{Rij05} for baryon-baryon in general.
Therefore, they apply also to hyperon-nucleon and we can refer for that
part of the potential to paper I.  
Also in the ESC-model, the potentials are of such a form that they are exactly 
equivalent in both momentum space and configuration space. 
The treatment of the mass differences among the baryons are handled exactly 
similar as is done in \cite{MRS89,RSY99}. Also, exchange potentials related to
strange meson exchanhe $K, K^*$ etc. , can be found in these references. 

The baryon mass differences in the intermediate states for TME- and MPE-
potentials has been neglected for {\it YN}-scattering. This, although possible
in principle, becomes rather laborious and is not expected to change the 
characteristics of the baryon-baryon potentials.

\subsection{Potentials and SU(3) Symmetry}
\label{sec:channels}
We consider all possible baryon-baryon interaction channels, where
the baryons are the members of the $J^P={\textstyle\frac{1}{2}}^+$
baryon octet
\begin{equation}
  B = \left( \begin{array}{ccc}
      {\displaystyle\frac{\Sigma^{0}}{\sqrt{2}}+\frac{\Lambda}{\sqrt{6}}}
               &  \Sigma^{+}  &  p  \\[2mm]
      \Sigma^{-} & {\displaystyle-\frac{\Sigma^{0}}{\sqrt{2}}
                   +\frac{\Lambda}{\sqrt{6}}}  &  n \\[2mm]
      -\Xi^{-} & \Xi^{0} &  {\displaystyle-\frac{2\Lambda}{\sqrt{6}}}
             \end{array} \right).
\label{eq:3.1}\end{equation}
The baryon masses, used in this paper, are given in Table~\ref{tabBmass}.
The meson nonets can be written as
\begin{equation}
     P=P_{\rm sin}+P_{\rm oct},
\label{eq:3.2}\end{equation}
where the singlet matrix $P_{\rm sin}$ has elements $\eta_0/\sqrt{3}$
on the diagonal, and the octet matrix $P_{\rm oct}$ is given by
\begin{equation}
   P_{\rm oct} = \left( \begin{array}{ccc}
      {\displaystyle\frac{\pi^{0}}{\sqrt{2}}+\frac{\eta_{8}}{\sqrt{6}}}
             & \pi^{+}  &  K^{+}  \\[2mm]
      \pi^{-} & {\displaystyle-\frac{\pi^{0}}{\sqrt{2}}
         +\frac{\eta_{8}}{\sqrt{6}}}  &   K^{0} \\[2mm]
      K^{-}  &  \overline{{K}^{0}}
             &  {\displaystyle-\frac{2\eta_{8}}{\sqrt{6}}}
             \end{array} \right),
\label{eq:3.3}\end{equation}
and where we took the pseudoscalar mesons with $J^P=0^+$ as a specific
example. Introducing the following notation for the isodoublets,
\begin{eqnarray}
  N &=& \left(\begin{array}{c} p \\ n \end{array} \right), \ \ \
  \Xi=\left(\begin{array}{c} \Xi^{0} \\ \Xi^{-} \end{array} \right), \ \ \
 {\rm and}\ \nonumber\\ 
  K &=& \left(\begin{array}{c} K^{+} \\ K^{0} \end{array} \right),
  \ \ \   K_{c}=\left(\begin{array}{c} \overline{K^{0}} \\
               -K^{-} \end{array} \right),        
\label{eq:3.4}\end{eqnarray}
the most general, SU(3) invariant, interaction Hamiltonian is then
given by~\cite{Swa63}
\begin{widetext}
\begin{eqnarray}
   {\cal H}_{\rm pv}^{\rm oct} &=&
  g_{N\!N\pi}(\overline{N}\bm{\tau}N)\!\cdot\!\bm{\pi}
  -ig_{\Sigma\Sigma\pi}(\overline{\bm{\Sigma}}\!\times\!\bm{\Sigma})
      \!\cdot\!\bm{\pi}
  +g_{\Lambda\Sigma\pi}(\overline{\Lambda}\bm{\Sigma}+
      \overline{\bm{\Sigma}}\Lambda)\!\cdot\!\bm{\pi}
  +g_{\Xi\Xi\pi}(\overline{\Xi}\bm{\tau}\Xi)\!\cdot\!\bm{\pi}
         +  \nonumber\\
 &&g_{\Lambda N\!K}\left[(\overline{N}K)\Lambda
         +\overline{\Lambda}(\overline{K}N)\right]
   +g_{\Xi\Lambda K}\left[(\overline{\Xi}K_{c})\Lambda
         +\overline{\Lambda}(\overline{K_{c}}\Xi)\right] + \nonumber\\
 &&g_{\Sigma N\!K}\left[\overline{\bm{\Sigma}}\!\cdot\!
         (\overline{K}\bm{\tau}N)+(\overline{N}\bm{\tau}K)
         \!\cdot\!\bm{\Sigma}\right]
   +g_{\Xi\Sigma K}\left[\overline{\bm{\Sigma}}\!\cdot\!
       (\overline{K_{c}}\bm{\tau}\Xi)
     +(\overline{\Xi}\bm{\tau}K_{c})\!\cdot\!\bm{\Sigma}\right]
                                +          \nonumber\\
 &&g_{N\!N\eta_{8}}(\overline{N}N)\eta_{8}
   +g_{\Lambda\Lambda\eta_{8}}(\overline{\Lambda}\Lambda)\eta_{8}
   +g_{\Sigma\Sigma\eta_{8}}(\overline{\bm{\Sigma}}\!\cdot\!
       \bm{\Sigma})\eta_{8}
   +g_{\Xi\Xi\eta_{8}}(\overline{\Xi}\Xi)\eta_{8}  +    \nonumber\\
 &&g_{N\!N\eta_{0}}(\overline{N}N)\eta_{0}
   +g_{\Lambda\Lambda\eta_{0}}(\overline{\Lambda}\Lambda)\eta_{0}
   +g_{\Sigma\Sigma\eta_{0}}(\overline{\bm{\Sigma}}\!\cdot\!
       \bm{\Sigma})\eta_{0}
   +g_{\Xi\Xi\eta_{0}}(\overline{\Xi}\Xi)\eta_{0},  
\label{eq:3.5}\end{eqnarray}
\end{widetext}
where we again took the pseudoscalar mesons as an example, dropped
the Lorentz character of the interaction vertices,
and introduced the charged-pion mass to make the pseudovector coupling
constant $f$ dimensionless.
All coupling constants can be expressed in terms of only four parameters.
The explicit expressions can be found in Ref.~\cite{RSY99}.
The $\Sigma$-hyperon is an isovector with phase chosen such~\cite{Swa63}
that
\begin{equation}
  \bm{\Sigma}\!\cdot\!\bm{\pi} = \Sigma^{+}\pi^{-}
       +\Sigma^{0}\pi^{0}+\Sigma^{-}\pi^{+}.
\label{eq:3.6}\end{equation}
This definition for $\Sigma^+$ differs from the standard Condon and
Shortley phase convention~\cite{Con35} by a minus sign. This means that,
in working out the isospin multiplet for each coupling constant in
Eq.~(\ref{eq:3.5}), each $\Sigma^+$ entering or leaving an interaction
vertex has to be assigned an extra minus sign. However, if the potential
is first evaluated on the isospin basis and then, via an isospin
rotation, transformed to the potential on the physical particle basis
(see below), this extra minus sign will be automatically accounted for.

In appendix~\ref{app:C}, Table~\ref{tab.irrep1} and Table~\ref{tab.irrep2}
we give the relation between the potentials on the isospin-basis, see
(\ref{eq:2.4})-(\ref{eq:2.5}), and the SU(3)-irreps.

Given the interaction Lagrangian (\ref{eq:3.5}) and a theoretical
scheme for deriving the potential representing a particular Feynman
diagram, it is now straightforward to derive the one-meson-exchange
baryon-baryon potentials.
We follow the Thompson approach~\cite{Tho70,Rij91,Rij92,Rij96} and
expressions for the potential in momentum space can be found in
Ref.~\cite{MRS89}. 
Since the nucleons have strangeness $S=0$, the hyperons $S=-1$,
and the cascades $S=-2$, the possible baryon-baryon interaction
channels can be classified according to their total strangeness,
ranging from $S=0$ for $N\!N$ to $S=-4$ for $\Xi\Xi$.
Apart from the wealth of accurate $N\!N$ scattering data for the total
strangeness $S=0$ sector, there are only a few, and not very accurate,
$Y\!N$ scattering data for the $S=-1$ sector, while there are no data
at all for the $S<-1$ sectors. We therefore believe that at this stage
it is not yet worthwhile to explicitly account for the small mass
differences between the specific charge states of the baryons and
mesons; i.e., we use average masses, isospin is a good quantum number,
and the potentials are calculated on the isospin basis. The possible
channels on the isospin basis are given in 
(\ref{eq:2.3}).

However, the Lippmann-Schwinger or Schr\"odinger equation is solved for
the physical particle channels, and so scattering observables are
calculated using the proper physical baryon masses.
The possible channels on the physical particle basis can be classified
according to the total charge $Q$; these are given in
(\ref{eq:2.2}).
The corresponding potentials are obtained
from the potential on the isospin basis by making the appropriate
isospin rotation. The matrix elements of the isospin rotation matrices
are nothing else but the Clebsch-Gordan coefficients for the two baryon
isospins making up the total isospin. (Note that this is the reason why
the potential on the particle basis, obtained from applying an isospin
rotation to the potential on the isospin basis, will have the correct
sign for any coupling constant on a vertex which involves a $\Sigma^+$.)

In order to construct the potentials on the isospin basis, we need
first the matrix elements of the various OBE exchanges between particular
isospin states. 
Using the iso-multiplets (\ref{eq:3.3}) and the Hamiltonian (\ref{eq:3.4})
the isospin factors can be calculated.
The results are given in Table~\ref{tab.iso},
where we use the pseudoscalar mesons as a specific example.
The entries contain the flavor-exchange operator $P_f$, which is
$+1$ for a flavor symmetric and $-1$ for a flavor anti-symmetric two-baryon
state. Since two-baryons states are totally anti-symmetric, one has
$P_f=-P_{x} P_{\sigma}$. Therefore, the exchange operator $P_f$ has the value
$P_f=+1$ for even-$L$ singlet and odd-$L$ triplet partial waves, and
$P_f=-1$ for odd-$L$ singlet and even-$L$ triplet partial waves.
In order to understand Table~\ref{tab.iso} fully, we have given in the 
following section Sec.~\ref{sec:exch} a general treatment of exchange forces.
This treatment shows also how to deal with the case where the initial/final
state involves identical particles and the final/inition state does not.

Second, we need to evaluate the TME and the MPE exchanges. The method
we used for these is the same as for hyperon-nucleon, and is described
in \cite{RY05}, Sec. IID.



\section{Exchange Forces}       
\label{sec:exch} 
The proper treatment of the flavor-exchange forces is for the $S < -2$-channels
more difficult than for the $S=0,1$-channels. The extra complication is the occurrence
of coupling between channels with identical and non-identical particles.
In order to understand the several $\sqrt{2}$-factors, see \cite{SR99}, we give 
here a systematic treatment of the flavor-exchange potentials.
The method followed is using a multi-channel framework, which starts starts by
ordering the two-particle states by assigning $A_i$ and $B_i$ for the channel
labeled with the index $i$, like in eq.~(\ref{eq:2.1}). The particles $A_i$ and
$B_i$ have CM-momenta $p_i$ and $p_i'$, spin components $s_i$ and $s_i'$.
The two-baryon states $|A_i B_i\rangle$ and $|B_i A_i\rangle$ are considered to
be distinct, leading to distinct two-baryon channels. The 'direct' and
the 'exchange' T-amplitudes are given by the T-matrix elements
\begin{equation}
 \langle A_j B_j| T_d|A_i B_i\rangle,\ \ \langle B_j A_j|T_e|A_iB_i\rangle\ ,
\label{exch.1}\end{equation}
and similarly for the direct and flavor-exchange potentials $V_d$ and $V_e$.
It is obvious from rotation invariance that 
\begin{eqnarray}
 \langle A_j B_j| T_d|A_i B_i\rangle &=& \langle B_j A_j|T_d|B_iA_i\rangle\ ,
 \nonumber\\ 
 \langle B_j A_j| T_e|A_i B_i\rangle &=& \langle A_j B_j|T_e|B_iA_i\rangle\ .    
\label{exch.2}\end{eqnarray}
A similar definition (\ref{exch.1}) and relation (\ref{exch.2}) apply for 
the direct and flavor-exchange potentials $V_d$ and $V_e$.

We notice that there is here no exchange of momenta or spin-components. So,
the momentum transfer for $V_d$ and $V_e$ is the same. Viewed from the coupled-channel
scheme this is the normal situation. 

\noindent The integral equations with two-baryon unitarity, e.g. 
the Thompson-, Lippmann-Schwinger-equation etc., reads for the $T_d$- and $T_e$-operator
\begin{widetext}
\begin{subequations}
\begin{eqnarray}
\langle A_jB_j|T_d|A_iB_i\rangle &=& \langle A_jB_j|V_d|A_iB_i\rangle + \sum_k \left[
\langle A_jB_j|V_d|A_kB_k\rangle\ G_k\ \langle A_kB_k|T_d|A_iB_i\rangle 
 \right.\nonumber\\ && \left. + 
\langle A_jB_j|V_e|A_kB_k\rangle\ G_k\ \langle A_kB_k|T_e|A_iB_i\rangle\ \right]\ , 
\\ 
\langle B_jA_j|T_e|A_iB_i\rangle &=& \langle B_jA_j|V_e|A_iB_i\rangle + \sum_k \left[
\langle B_jA_j|V_d|B_kA_k\rangle\ G_k\ \langle B_kA_k|T_e|A_iB_i\rangle 
 \right.\nonumber\\ && \left. + 
\langle B_jA_j|V_e|A_kB_k\rangle\ G_k\ \langle A_kB_k|T_d|A_iB_i\rangle\ \right]\ . 
\label{exch.3}\end{eqnarray}
\end{subequations}
These coupled equations can be diagonalized by introducing the T- and V-operators
\begin{equation}
 T^{\pm} = T_d \pm T_e\ ,\ \ V^{\pm} = V_d \pm V_e\ .    
\label{exch.4}\end{equation}
which satisfy separate integral equations
\begin{eqnarray}
\langle A_jB_j|T^\pm|A_iB_i\rangle &=& \langle A_jB_j|V^\pm|A_iB_i\rangle 
+ \sum_k
\langle A_jB_j|V^\pm|A_kB_k\rangle\ G_k\ \langle A_kB_k|T^\pm|A_iB_i\rangle\ .
\label{exch.5}\end{eqnarray}
Notice that on the basis of states with definite flavor symmetry
\begin{equation}
 |A_i B_i\rangle_{\pm} = \frac{1}{\sqrt{2}}\left[ |A_i B_i \rangle 
 \pm |B_i A_i\rangle\right]\ , 
\label{exch.6}\end{equation}
the $T^\pm$ and $V^\pm$ matrix elements are also given by
\begin{equation}
  T^\pm_{ij} = _\pm\!\!\langle A_i B_i | T | A_j B_j\rangle_\pm\ ,\ \  
  V^\pm_{ij} = _\pm\!\!\langle A_i B_i | V | A_j B_j\rangle_\pm\ .     
\label{exch.7}\end{equation}
\end{widetext}

\subsection{Identical Particles}
Sofar, we considered the general case where $A_i \neq B_i$ for all channels.
In the case that $A_i=B_i$ for some $i$, one has
$\langle B_iA_i|V_e  |A_iB_i\rangle = 0$, because there is no distinct 
physical state corresponding to the 'flavor exchange-state'. For example for
a flavor single channel like $pp$ one deduces from (\ref{exch.3}) that then also
$T_e=0$, and one has in this case the integral equation
\begin{eqnarray}
&& \langle A_jB_j|T_d|A_iB_i\rangle = \langle A_jB_j|V_d|A_iB_i\rangle +
 \nonumber\\ && \sum_k 
\langle A_jB_j|V_d|A_kB_k\rangle\ G_k\ \langle A_kB_k|T_d|A_iB_i\rangle\ , 
\label{exch.8}\end{eqnarray}
where the labels $i$ and $j$ now denote e.g. the spin-components.

\subsection{Coupled $\Lambda\Lambda$ and $\Xi N$ system}
This multi-channel system represents the case where there is mixture of channels with
identical and with non-identical particles. The three states we distinguish are
$|\Lambda\Lambda\rangle, |\Xi N\rangle$, and $|N\Xi\rangle$. Choosing the same 
ordering, the potential written as a $3x3$-matrix reads
\begin{equation}
V = \left(\begin{array}{ccc} 
 (\Lambda\Lambda|V|\Lambda\Lambda) & (\Lambda\Lambda|V| \Xi N) & 
 (\Lambda\Lambda|V| N \Xi \\
 (\Xi N |V|\Lambda\Lambda) & (\Xi N |V| \Xi N) & (\Xi N |V| N \Xi \\
 (N \Xi |V|\Lambda\Lambda) & (N \Xi |V| \Xi N) & (N \Xi |V| N \Xi   
 \end{array}\right)\ .
\label{exch.9}\end{equation}
With a similar notation for the T-matrix, 
the Lippmann-Schwinger equation can be written compactly as a $3x3$-matrix
equation:
\begin{equation}
   T = V + V\ G\ T\ ,\ {\rm with}\ G_{ij}= G_i\ \delta_{ij}\ .
\label{exch.10}\end{equation}
Next, we make a transformation to states, which are either symmetric or anti-symmetric
for particle interchange. Then, according to the discussion above, we can separate them
in the Lippmann-Schwinger equation. This is achieved by the transformation
\begin{widetext}
\begin{equation}
 \left(\begin{array}{c} 
 \Lambda\Lambda \\ \Xi N \\ N \Xi \end{array}\right) \Rightarrow
 \left(\begin{array}{c} 
 \Lambda\Lambda \\ (\Xi N + N \Xi)/\sqrt{2} \\ 
 (\Xi N - N \Xi)/\sqrt{2} \end{array}\right) =             
 \left(\begin{array}{ccc} 1 & 0 & 0 \\ 0 & 
 1/\sqrt{2} & 1/\sqrt{2} \\
 0 & 1/\sqrt{2} &-1/\sqrt{2}    
 \end{array}\right)
 \left(\begin{array}{c} 
 \Lambda\Lambda \\ \Xi N \\ N \Xi \end{array}\right)\ .             
\label{exch.11}\end{equation}
one gets in the transformed basis for the potential
\begin{equation}
 UVU^{-1} = \left(\begin{array}{cll} 
 V_{\Lambda\Lambda;\Lambda\Lambda} & 
 (V_{\Lambda\Lambda;\Xi N} + V_{\Lambda\Lambda;N \Xi})/\sqrt{2} & 
 (V_{\Lambda\Lambda;\Xi N} - V_{\Lambda\Lambda;N \Xi})/\sqrt{2} \\
 (V_{\Xi N;\Lambda\Lambda} + V_{N \Xi;\Lambda\Lambda})/\sqrt{2} & 
 (V_{\Xi N;\Xi N} + V_{\Xi N; N \Xi}) & \hspace{1.43cm} 0 \\
 (V_{\Xi N;\Lambda\Lambda} - V_{N \Xi;\Lambda\Lambda})/\sqrt{2} &\hspace{1.43cm}  0 &
 (V_{\Xi N;\Xi N} - V_{\Xi N; N \Xi}) \\
\end{array}\right)\ ,
\label{exch.12}\end{equation}
and of course, a similar form is obtained for the T-matrix on the transformed basis.
Now, obviously we have that $V_{\Lambda\Lambda;\Xi N} = V_{\Lambda\Lambda;N \Xi}$ and          
$V_{\Xi N;\Lambda\Lambda} = V_{N \Xi;\Lambda\Lambda}$. Therefore, one sees that
the even and odd states under particle exchange are decoupled in (\ref{exch.12}).
Also $V_{\Xi N;\Lambda\Lambda} + V_{N \Xi;\Lambda\Lambda}= \sqrt{2}
V_{\Xi N;\Lambda\Lambda}$, etc. showing the           
appearance of the $\sqrt{2}$-factors, mentioned before. Indeed, they appear
in a systematic way using the multi-channel framework.

\subsection{The K-exchange Potentials}                         
Consider for example the $(\Lambda\Lambda, \Xi N)$-system, having $I=0$.
Mesons with strangeness, $K(495)$, $K^*(892)$, $\kappa(900)$, $K_1(1270)$, are obviously 
the only ones that can give transition potentials, i.e. 
$ V_{\Lambda\Lambda;\Xi N} \neq 0$ and $ V_{\Xi N;\Lambda\Lambda} \neq 0$.  
The $\Xi N(I=0)$-states anti-symmetric and
symmetric in flavor are respectively:
\begin{subequations}
\begin{eqnarray}
&& P_f=-1:\ \frac{1}{\sqrt{2}}\left[|\Xi N(I=0)\rangle - |N \Xi(I=0)\rangle\right]\ ,
\\ 
&& P_f=+1:\ \frac{1}{\sqrt{2}}\left[|\Xi N(I=0)\rangle + |N \Xi(I=0)\rangle\right]\ .
\label{exch.13}\end{eqnarray}
\end{subequations}
Analyzing the $^1S_0$-state one has because of the anti-symmetry of the two-fermion state
w.r.t. the exchange of all quantum labels, $P_f = -P_\sigma P_x = +1$, where
$P_f$ denotes the flavor-symmetry.  
Taking here the $K(495)$ as a generic example, and using (\ref{eq:3.4}) and (\ref{eq:3.5}),
one finds that
\begin{equation}
\langle \Xi^0 n|V(K)|\Lambda\Lambda\rangle = + g_{K\Lambda N} g_{K\Xi N},\ \
\langle \Xi^- p|V(K)|\Lambda\Lambda\rangle = - g_{K\Lambda N} g_{K\Xi N}\ .  
\label{exch.14}\end{equation}
Taking here the $K(495)$ as a generic example, 
Then, since
$|\Xi N(I=0)\rangle = \left[|\Xi^0 n\rangle - |\Xi^- p\rangle\right]/\sqrt{2}$, 
one obtains for the 'direct' potential the coupling
\begin{equation}
V_{\Xi N;\Lambda\Lambda} = \langle \Xi N(I=0)\rangle | V_d(K)|
\Lambda\Lambda\rangle \Leftarrow \sqrt{2} g_{K\Lambda N} g_{K\Xi N}\ .
\label{exch.15}\end{equation}
The same result is found for the 'exchange' potential $V_{N\Xi;\Lambda\Lambda}$. 
Therefore
\begin{equation}
\frac{1}{\sqrt{2}}\left(V_{\Xi N;\Lambda\Lambda} + V_{N \Xi;\Lambda\Lambda} 
\right) = \langle \Xi N(I=0) | V(K)|
\Lambda\Lambda\rangle \Leftarrow 2 g_{K\Lambda N} g_{K\Xi N}\ ,
\label{exch.16}\end{equation}
which has indeed the $(1+P_f)$-factor given in Table~\ref{tab.iso},  
and is identical to Table IV in \cite{SR99}, for
$(\Lambda\Lambda|K|\Xi N)$.                  

For 
$\langle\Xi N|K|\Sigma\Lambda\rangle$ the entry for $I=1$ consists of two parts. These 
correspond to $V_d \propto g_{\Lambda N K} g_{\Xi\Sigma K}$ and  
$V_e \propto g_{\Sigma N K} g_{\Xi\Lambda K}$ respectively, i.e. the direct and      
exchange contributions involve different couplings. Therefore, they are not
added together.

\subsection{The $\eta$- and $\pi$-exchange Potentials}                         
Next, we discuss briefly the computation of the 
entries for $\eta$- and $\pi$-exchange in Table~\ref{tab.iso}. 
First, the entries with --- indicate that the corresponding physical state does
not exist. Next we give further specific remarks and calculations:
\begin{enumerate}
\item[a.] For $\eta,\eta'$-exchange one has that $V_e=0$. The matrix elements for the
$\Lambda\Lambda$- and $\Xi N$-state are easily seen to be correct. For the 
$\Sigma\Sigma$-states one has $P_f=1$ for $I_{\Sigma\Sigma}=0,2$, and 
$P_f=-1$ for $I_{\Sigma\Sigma}=1$. This explains the $\Sigma\Sigma$ matrix 
element.
\item[b.] For $\langle\Xi N|\pi|\Xi N\rangle$ the computation is identical to that 
for NN, in particular $p n$.
\item[c.] For $\langle\Sigma\Sigma|\pi|\Sigma\Sigma\rangle$ consider the $I=0,I_3=0$ and
$I=1,I_3=0$ matrix elements. In these cases one has $V_e=0$ as one can easily check.
Then, using the cartesian base, we have for 
$\langle\Sigma_i\Sigma_m|\pi|\Sigma_j\Sigma_n\rangle \Rightarrow
-g_{\Sigma\Sigma\pi}^2 \sum_{p=1}^3 \epsilon_{jip}\epsilon_{nmp}=
 -g_{\Sigma\Sigma\pi}^2 (\delta_{jn}\delta_{im}-\delta_{jm}\delta_{in})$. 
Employing the states $|I=0,I_3=0\rangle \sim -\sum_{i,m=1}^3\delta_{im}|\Sigma_i\Sigma_m\rangle
/\sqrt{3}$ and 
$|I=1,I_3=0\rangle \sim -i\sum_{i,m=1}^3\epsilon_{im3}|\Sigma_i\Sigma_m\rangle/\sqrt{2}$,
one obtains the results in Table~\ref{tab.iso}.
\end{enumerate}
With the ingredients given above one can easily check the other entries in Table~\ref{tab.iso}.
 
\vspace{3mm}

\end{widetext}
 
\begin{table}
\caption{Isospin factors for the various meson exchanges in the
         different total strangeness and isospin channels.
         $P_f$ is the flavor-exchange operator.
         The $I=2$ case only contributes to $S=-2$ $\Sigma\Sigma$
         scattering, where the isospin factors can collectively be
         given by $(\Sigma\Sigma|\eta,\eta',\pi|\Sigma\Sigma)=
         {\protect\textstyle\frac{1}{2}}(1+P_f)$, and so they are
         not separately displayed in the table.
         Non-exixisting channels are marked by a long-dash.
         }
\begin{ruledtabular}
\begin{tabular}{ccc} 
  $S=-2$ & $I=0$ & $I=1$ \\
\colrule 
  $(\Lambda\Lambda|\eta,\eta'|\Lambda\Lambda)$
                       & ${\textstyle\frac{1}{2}}(1+P_f)$ & --- \\[.5mm]
  $(\Xi N|\eta,\eta'|\Xi N)$ & ${\textstyle\frac{1}{2}}(1+P_f)$ & 1 \\[.5mm]
  $(\Sigma\Sigma|\eta,\eta'|\Sigma\Sigma)$
                       & ${\textstyle\frac{1}{2}}(1+P_f)$
                       & ${\textstyle\frac{1}{2}}(1-P_f)$ \\
  $(\Sigma\Lambda|\eta,\eta'|\Sigma\Lambda)$ & --- & 1 \\
  $(\Xi N|\pi|\Xi N)$  & $-3$ & 1 \\
  $(\Sigma\Sigma|\pi|\Sigma\Sigma)$ & $-(1+P_f)$
                       & $-{\textstyle\frac{1}{2}}(1-P_f)$ \\[.5mm]
  $(\Lambda\Lambda|\pi|\Sigma\Sigma)$
                       & $-{\textstyle\frac{1}{2}}\sqrt{3}(1+P_f)$ & --- \\
  $(\Sigma\Lambda|\pi|\Lambda\Sigma)$ & --- & $P_f$ \\
  $(\Sigma\Sigma|\pi|\Sigma\Lambda)$ & --- & $(1-P_f)$ \\
  $(\Lambda\Lambda|K|\Xi N)$ & $1+P_f$ & --- \\
  $(\Sigma\Sigma|K|\Xi N)$   & $\sqrt{3}(1+P_f)$ & $\sqrt{2}(1-P_f)$ \\
  $(\Xi N|K|\Sigma\Lambda)$  & --- & $\sqrt{2};-P_f\sqrt{2}$ \\
\end{tabular}
\end{ruledtabular}
\label{tab.iso}
\end{table}

\section{Multi-channel Thresholds and Potentials}
\label{sec:thresh}
\subsection{Thresholds}
Clearly, the $S=-2$ two-baryon channels represent a number of separate 
coupled-channel systems, separated by the charge, see (\ref{eq:2.2}).
A further subdivision is according to the total isospin. 
The different thresholds have been discussed in detail in \cite{SR99}, and 
we show them here in Fig.~\ref{thresh1} for  the purpose of 
general orientation. Their presence turns the
Lippmann-Schwinger and Schr\"odinger equation into a coupled-channel
matrix equation, where the different channels open up at different energies.
In general one has a combination of 'open' and 'closed' channels. For a 
discussion of the solution of such a mixed system, we refer to \cite{Swa71}.

\subsection{Threshold- and Meson-mass corrections in Potentials}
As discussed in \cite{SR99}, the one-meson-exchange Feynman-graph consists
actually of two three-dimensional time-ordered graphs. 
The energy denominator from these two diagrams reads
\begin{equation}
   D(\omega)=\frac{1}{2\omega}\left[\frac{1}{E_2+E_3-W+\omega}
             +\frac{1}{E_1+E_4-W+\omega}\right],
\label{4.1} \end{equation}
where, $W=\sqrt{s}$ is the total energy and $\omega^2={\bf k}^2+m^2$,
with $m$ the meson mass and ${\bf k}={\bf p}'-{\bf p}$ the momentum
transfer. 
From (\ref{4.1}) it is clear that the potential is energy dependent.
We use the static approximation $E_i\rightarrow M_i$ and
$W\rightarrow M^0_1+M^0_2$, where the superscript 0 refers to the masses of the 
lowest threshold of the particular coupled-channel system q, see (\ref{eq:2.2}). 
They are in general not 
equal to the masses $M_1$ and $M_2$ occurring in the time-ordered
diagrams. For example, the potential for the $\Sigma\Sigma$
contribution in the coupled-channel $\Lambda\Lambda$ system has
$M_1=M_2=M_{\Sigma}$, but $M^0_1=M^0_2=M_{\Lambda}$.
Denoting $a \equiv E_2+E_3-W \approx M_2+M_3-M^0_1-M^0_2>0$, and similarly
for $E_1+E_4-W$, we have for the 'propagators' \cite{Rij92} for $0<a<m$
\begin{equation}
  \frac{1}{\omega(\omega+a)} =\frac{2}{\pi}\int_0^{\infty}
          \frac{ad\lambda}{(a^{2}+\lambda^2)(\omega^2+\lambda^2)}\ .
\label{4.2} \end{equation}
This integral representation makes it possible to deal with it numerically
rather exactly. However, we think that such a sophistication is unnecessary 
at present nor for a description of the $S=-1$ scattering data, nor for $S=-2$.
where there are virtually no data at all.
Therefore, we handle with this approximately as follows:\\

\noindent 1.\ \underline{Elastic potentials}: In this case we use (\ref{4.2}), and 
in (\ref{4.1}) one has $E_1=E_3 \approx M_i$ and $E_2=E_4 \approx M'_i$, for
the elastic channel, label $i$. Here $a \approx M_i+M'_i -M_1^0-M_2^0$. Then,  
\begin{eqnarray}
   D_i(\omega)&=&\frac{1}{\omega^2} + \Delta_i(\omega,a),\ \  
   \Delta_i(\omega,a) = \frac{2}{\pi}\int_0^{\infty}
   \frac{d\lambda}{a^2+\lambda^2}\cdot \nonumber \\ && \times
 \left[ \frac{1}{\omega^2}-\frac{1}{\omega^{2}+\lambda^2}\right]\ ,
\label{4.3} \end{eqnarray}
for $0<a<m$. Because of this condition we apply this not to the pseudoscalars,
but to the vector-, scalar-, and axial-mesons. For example, in the case of the
$\Lambda\Lambda$-scattering, the $\Sigma\Sigma$-channel potential is reduced
by this effect. Since the $\Sigma\Sigma$-channel is rather far away from the
others, we in practice apply (\ref{4.3}) only for that channel. In this case $a>0$
and the $\theta(-a)$-term in $D_i(\omega)$ vanishes.\\

\noindent 2.\ \underline{Inelastic potentials}: In this case, like in \cite{SR99} and all
other papers on the Nijmegen potentials, we use the approximation of \cite{Swa62},
using the fact that $M_1^0+M_2^0$ is mostly rather close to the average of the
initial and final-sate baryon masses. Then, the propagator can be written as
\begin{equation}
   D(\omega)\rightarrow \frac{1}{\omega^2-{\textstyle\frac{1}{4}}
            (M_3-M_4+M_2-M_1)^2},      
\label{4.4} \end{equation}
which amounts to introducing an effective meson mass $\overline{m}$
\begin{equation}
    m^2 \rightarrow \overline{m}^2=m^2-{\textstyle\frac{1}{4}}
                 (M_3-M_4+M_2-M_1)^2.
\label{4.5} \end{equation}
For more details of this effect on the exchanged meson masses, we refer to 
\cite{SR99}.

\begin{table}
\caption{Baryon masses in MeV/$c^2$.}
\begin{ruledtabular}
\begin{tabular}{lcr}
  Baryon & & Mass \\
\hline      
  Nucleon & $p$            &  938.2796  \\
          & $n$            &  939.5731  \\
  Hyperon & $\Lambda$      & 1115.60    \\
          & $\Sigma^+$     & 1189.37    \\
          & $\Sigma^0$     & 1192.46    \\
          & $\Sigma^-$     & 1197.436   \\
  Cascade & $\Xi^0$        & 1314.90    \\
          & $\Xi^-$        & 1321.32    \\
          &                &            \\
\hline      
\end{tabular}
\end{ruledtabular}
\label{tabBmass}
\end{table}
The used baryon masses are about the same as in \cite{SR99}, and are given in
Table~\ref{tabBmass}.
The used meson masses are the same as in paper II \cite{RY05}, as well as the 
cut-off mases.

\section{Results}
\label{sec:results}
The main purpose of this paper is to present the properties of the
four ESC04 potentials for the $S=-2$ sector. We will show the
detailed results for ESC04a and ESC04d, which are sufficient to
represent the possible kind of results. Model ESC04a is representative
for the inclusion of SU(3)-breaking of the couplings, and ESC04d 
for the case of SU(3)-symmetric couplings.
The free parameters in each model are fitted to the $N\!N$ and $Y\!N$
scattering data for the $S=0$ and $S=-1$ sectors, respectively.
Given the expressions for the coupling constants in terms of the octet
and singlet parameters and their values for the six different models
as presented in Ref.~\cite{RSY99}, it is straightforward to evaluate
all possible baryon-baryon-meson coupling constants needed for the
$S\leq-2$ potentials. A complete set of coupling constants for models
ESC04a and ESC04d is given in Tables~\ref{tabcop04a} and \ref{tabcop04d},
respectively. 

\begin{figure}   
  \resizebox{3.5cm}{!}
 {\includegraphics[200, 50][400,750]{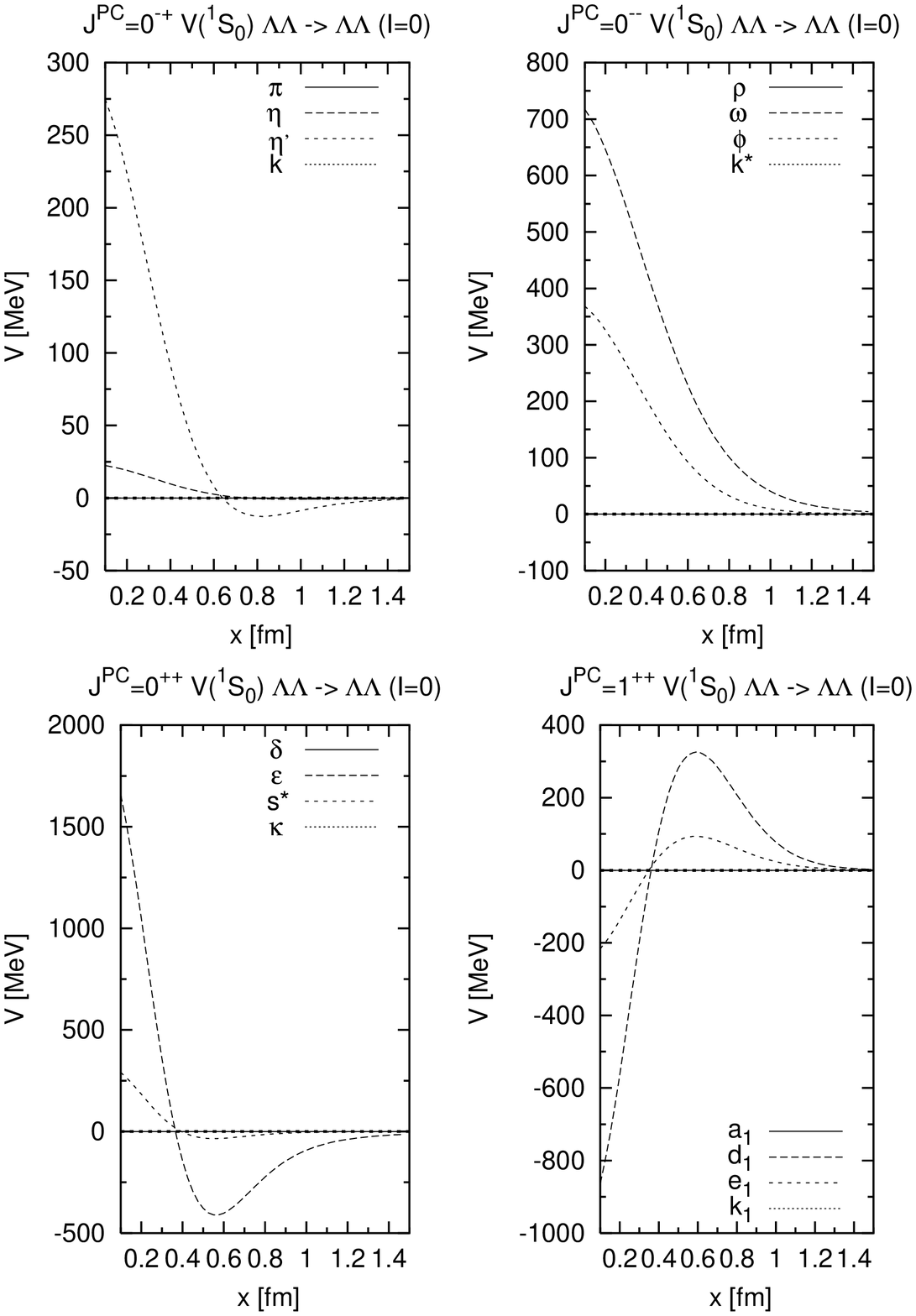}}
\caption{ESC04d: OBE contributions to the $(^1S_0, I=0)$ potentials                 
 for the PS, V, S, and A meson nonets.}
\label{obefig1}
 \end{figure}

\begin{figure}   
  \resizebox{3.5cm}{!}
 {\includegraphics[200, 50][400,750]{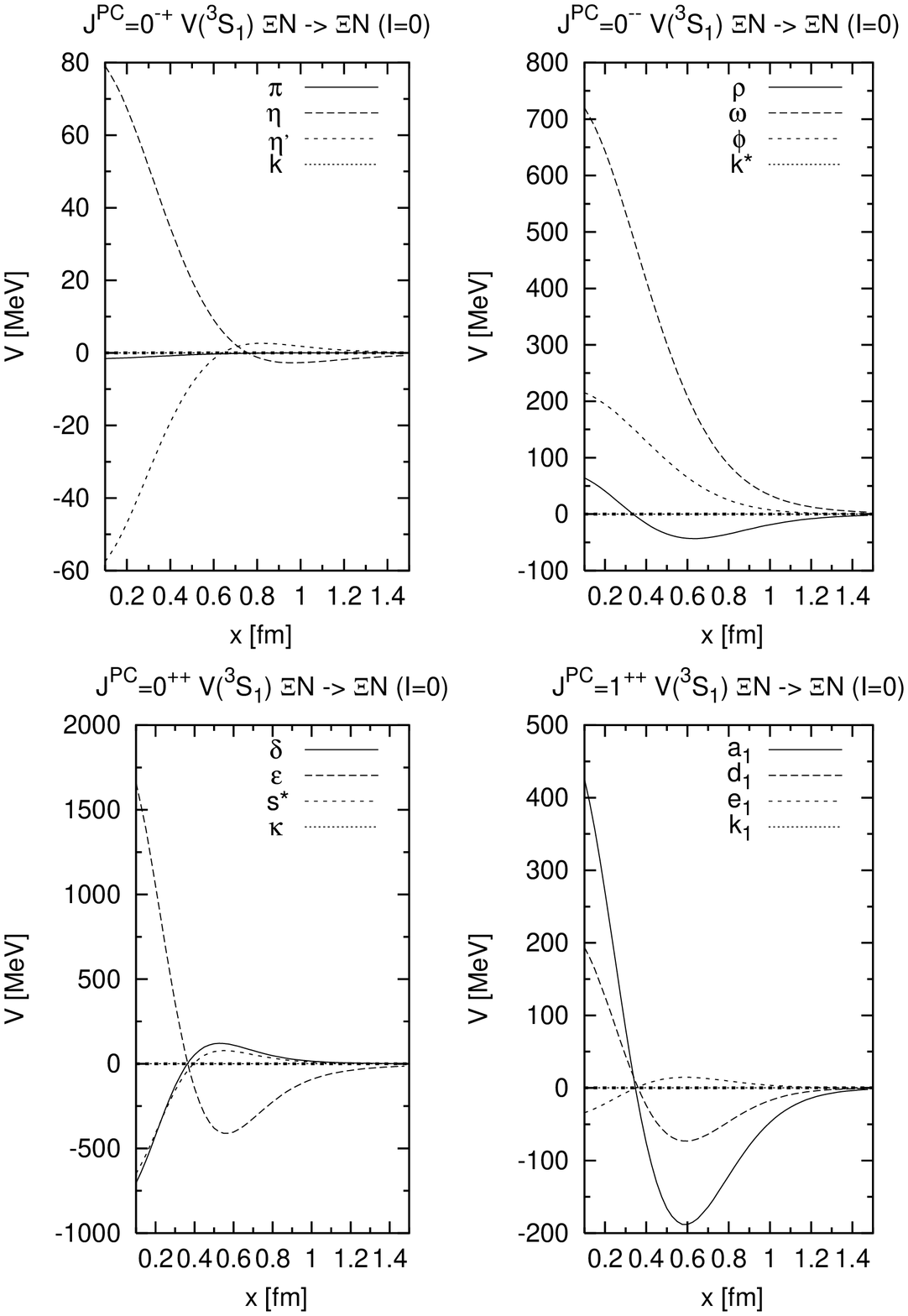}}
\caption{ESC04d: OBE contributions to the $(^3S_1, I=0)$ potentials                 
 for the PS, V, S, and A meson nonets.}
 \label{obefig2}
 \end{figure}

In Fig's~\ref{obefig1} and Fig.~\ref{obefig2} we display the OBE potentials
for the individual pseudoscalar, vector, scalar, and axial mesons in the
case of model ESC04d.

\begin{widetext}

\begin{table}
\caption{Coupling constants for model ESC04a, divided by
         $\protect\sqrt{4\pi}$. $M$ refers to the meson.
         The coupling constants are listed in the order pseudoscalar,
         vector ($g$ and $f$), scalar, and diffractive.}
\begin{ruledtabular}
\begin{tabular}{ccrrrrccrrrr}
 Type & $M$ & $N\!N\!M$ & $\Sigma\Sigma M$ & $\Sigma\Lambda M$ & $\Xi\Xi M$
 & $\hspace*{2ex}$
 & $M$ & $\Lambda N\!M$ & $\Lambda\Xi M$ & $\Sigma N\!M$ & $\Sigma\Xi M$ \\
\colrule
 $f$ & $\pi$    &   0.2631  &   0.2456  &   0.1620  & --0.0175 
    && $K$      & --0.2179  &   0.0977  &   0.0130  & --0.1951  \\
 $g$ & $\rho$   &   0.7800  &   1.5600  &   0.0000  &   0.7800 
    && $K^*$    & --1.0022  &   1.0022  & --0.5786  & --0.5786  \\
 $f$ &          &   3.4711  &   1.9177  &   2.9009  & --1.5535 
    &&          & --2.3080  &   0.1560  &   1.1524  & --2.5750  \\
 $g$ & $a_1$    &   2.5426  &   1.1922  &   2.2477  & --1.3505 
    && $K_1$    & --1.1597  & --0.0492  &   0.7263  & --1.3675  \\
 $g$ & $a_0$    &   0.9251  &   1.5562  &   0.1698  &   0.6311 
    && $\kappa$ & --1.0506  &   0.9261  & --0.4628  & --0.6785  \\
 $g$ & $a_2$    &   0.00000 &   0.00000 &   0.00000 &   0.00000
    && $K^{**}$ &   0.00000 &   0.00000 &   0.00000 &   0.00000 \\[2mm]
 Type & $M$ & $N\!N\!M$ & $\Lambda\Lambda M$ & $\Sigma\Sigma M$ & $\Xi\Xi M$
 & $\hspace*{2ex}$
 & $M$ & $N\!N\!M$ & $\Lambda\Lambda M$ & $\Sigma\Sigma M$ & $\Xi\Xi M$ \\
\colrule
 $f$ & $\eta$        &   0.1933  & --0.0203  &   0.2153  & --0.1161 
    && $\eta'$       &   0.1191  &   0.1421  &   0.1167  &   0.1525  \\
 $g$ & $\omega$      &   3.0135  &   2.2104  &   2.2104  &   1.4073 
    && $\phi$        & --0.3849  & --0.9611  & --0.9611  & --1.5372  \\
 $f$ &               &   0.4467  & --1.4028  &   2.0461  & --1.5278 
    &&               & --0.0502  & --1.3771  &   1.0973  & --1.4667  \\
 $g$ & $f_1$         &   1.0190  &   0.7973  &   1.2595  &   0.8067 
    && $f'_1$        &   1.0352  & --0.2915  &   2.4744  & --0.2352  \\
 $g$ & $\varepsilon$ &   3.4635  &   2.5842  &   2.7926  &   1.8090 
    && $f_0$         & --0.8162  & --1.3699  & --1.2387  & --1.8580  \\
 $g$ & $P$           &   1.9651  &   1.9651  &   1.9651  &   1.9651 
    && $f_2$         &   0.0000  &   0.0000  &   0.0000  &   0.0000 
\end{tabular}
\end{ruledtabular}
\label{tabcop04a}
\end{table}

\begin{table}
\caption{Coupling constants for model ESC04d, divided by
         $\protect\sqrt{4\pi}$. $M$ refers to the meson.
         The coupling constants are listed in the order pseudoscalar,
         vector ($g$ and $f$), scalar, and diffractive.}
\begin{ruledtabular}
\begin{tabular}{ccrrrrccrrrr}
 Type & $M$ & $N\!N\!M$ & $\Sigma\Sigma M$ & $\Sigma\Lambda M$ & $\Xi\Xi M$
 & $\hspace*{2ex}$
 & $M$ & $\Lambda N\!M$ & $\Lambda\Xi M$ & $\Sigma N\!M$ & $\Sigma\Xi M$ \\
\colrule
 $f$ & $\pi$    &   0.2599  &   0.2592  &   0.1505  & --0.0008 
    && $K$      & --0.2997  &   0.1492  &   0.0008  & --0.2599  \\
 $g$ & $\rho$   &   0.7038  &   1.4076  &   0.0000  &   0.7038 
    && $K^*$    & --1.2190  &   1.2190  & --0.7038  & --0.7038  \\
 $f$ &          &   3.2909  &   2.8332  &   2.1642  & --0.4577 
    &&          & --3.5357  &   1.3715  &   0.4577  & --3.2909  \\
 $g$ & $a_1$    &   2.4310  &   1.1398  &   2.1490  & --1.2912 
    && $K_1$    & --2.0616  & --0.0874  &   1.2912  & --2.4310  \\
 $g$ & $a_0$    &   1.0303  &   1.7331  &   0.1891  &   0.7028 
    && $\kappa$ & --1.5955  &   1.4064  & --0.7028  & --1.0303  \\
 $g$ & $a_2$    &   0.00000 &   0.00000 &   0.00000 &   0.00000
    && $K^{**}$ &   0.00000 &   0.00000 &   0.00000 &   0.00000 \\[2mm]
 Type & $M$ & $N\!N\!M$ & $\Lambda\Lambda M$ & $\Sigma\Sigma M$ & $\Xi\Xi M$
 & $\hspace*{2ex}$
 & $M$ & $N\!N\!M$ & $\Lambda\Lambda M$ & $\Sigma\Sigma M$ & $\Xi\Xi M$ \\
\colrule
 $f$ & $\eta$        &   0.2125  & --0.0634  &   0.2137  & --0.2007 
    && $\eta'$       &   0.1188  &   0.2359  &   0.1183  &   0.2942  \\
 $g$ & $\omega$      &   3.0366  &   2.2944  &   2.2944  &   1.5523 
    && $\phi$        & --0.7935  & --1.7606  & --1.7606  & --2.7277  \\
 $f$ &               &   0.0052  & --2.1472  &   0.4878  & --2.9822 
    &&               &   1.7248  & --1.0803  &   2.3537  & --2.1684  \\
 $g$ & $f_1$         &   1.7228  &   2.5283  &   0.8489  &   2.4941 
    && $f'_1$        &   0.6363  & --1.2614  &   2.6950  & --1.1810  \\
 $g$ & $\varepsilon$ &   3.5434  &   3.2267  &   3.3017  &   2.9475 
    && $f_0$         &   0.7172  & --0.8465  & --0.4759  & --2.2249  \\
 $g$ & $P$           &   2.3532  &   2.3532  &   2.3532  &   2.3532 
    && $f_2$         &   0.0000  &   0.0000  &   0.0000  &   0.0000 
\end{tabular}
\end{ruledtabular}
\label{tabcop04d}
\end{table}

\end{widetext}

In the following we will present the model predictions for scattering
lengths, bound states, and cross sections. 

 \begin{figure}   
 \begin{center}
 \resizebox{9.5cm}{11.25cm} 
 {\includegraphics[000,175][400,625]{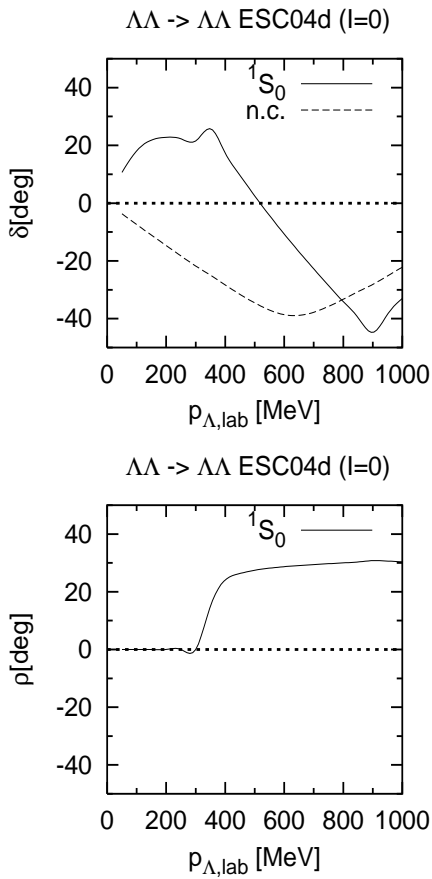}}
\caption{ESC04d $(^1S_0, I=0)$ $\Lambda\Lambda$-phases.
 The dashed curve n.c. is the case with no coupling to 
 the $\Xi N, \Sigma\Sigma$ channels.  }  
 \label{phasll0} 
 \end{center}
 \end{figure}

 \begin{figure}   
 \resizebox{7.0cm}{11.25cm} 
 {\includegraphics[100,175][500,625]{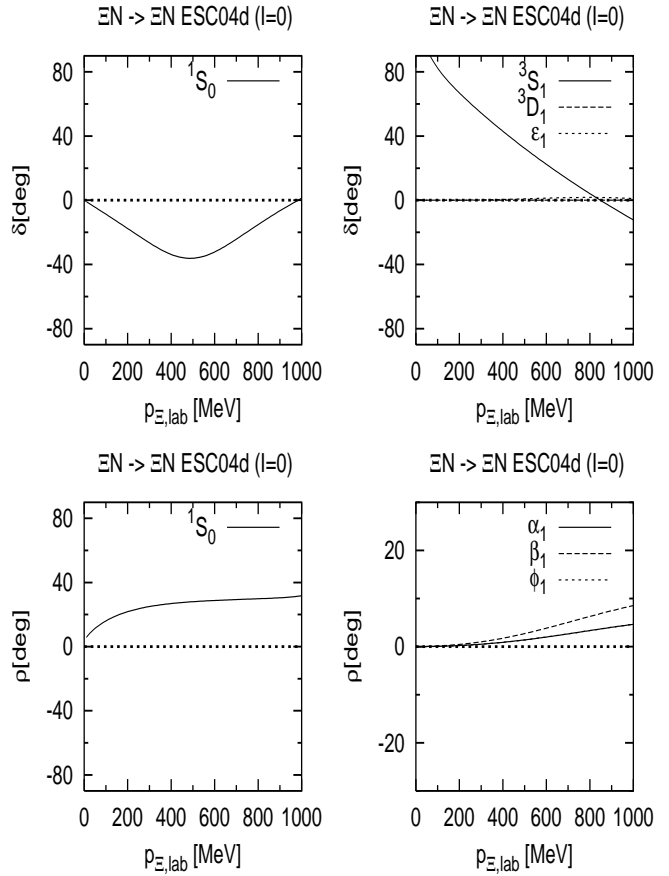}}
\caption{ESC04d $I=0$ $\Xi N$-phases.}  
 \label{phasxn0} 
 \end{figure}

 \begin{figure}   
 \resizebox{7.0cm}{11.25cm} 
 {\includegraphics[100,175][500,625]{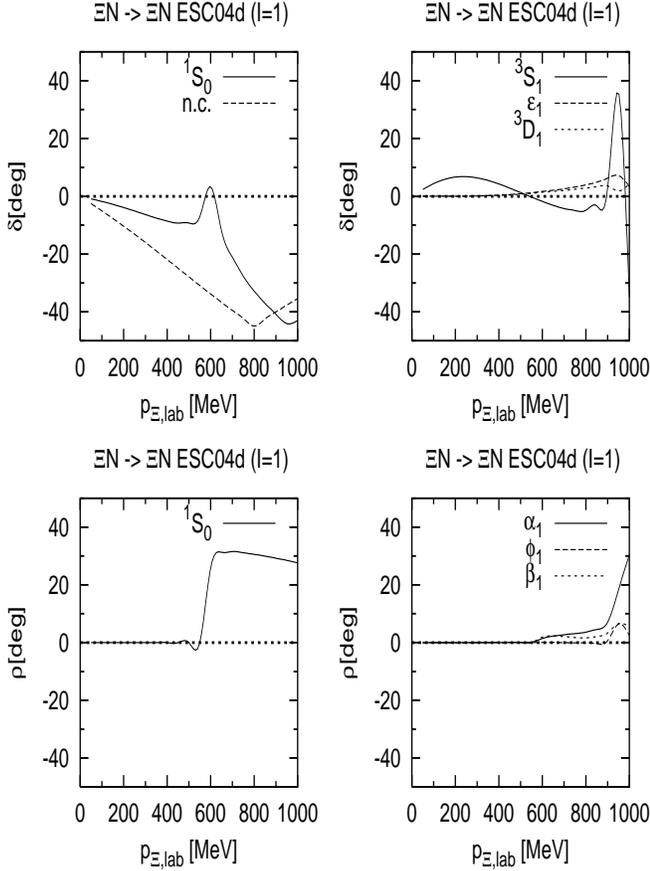}}
\caption{ESC04d $I=1$ $\Xi N$-phases.
 The dashed curve n.c. is the case with no coupling to 
 the $\Sigma\Lambda, \Sigma\Sigma$ channels.  }  
 \label{phasxn1} 
 \end{figure}

 \begin{figure}   
 \resizebox{7.0cm}{11.25cm} 
 {\includegraphics[100,175][500,625]{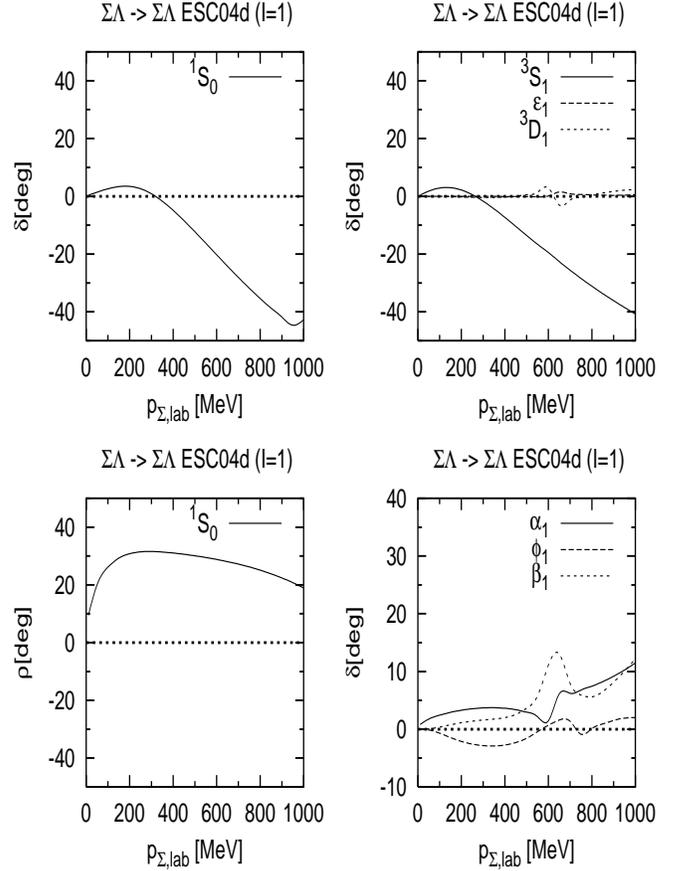}}
\caption{ESC04d $I=1$ $\Sigma\Lambda$-phases.}  
 \label{phassl1} 
 \end{figure}


\subsection{Effective-range parameters}

For ESC04a the $I=0$ low-energy parameters are
\begin{eqnarray*}
 a_{\Lambda\Lambda}(^1S_0) &=& -3.804\ [fm]\ ,\ \ 
 r_{\Lambda\Lambda}(^1S_0)  =   2.420\ [fm]\ .    
\label{le.1}\end{eqnarray*}
\begin{eqnarray*}
 a_{\Xi N}(^3S_1) &=& -1.672\ [fm]\ ,\ \ 
 r_{\Xi N}(^3S_1)  =   2.704\ [fm]\ .    
\label{le.2}\end{eqnarray*}

For ESC04d the $I=0$ low-energy parameters are
\begin{eqnarray*}
 a_{\Lambda\Lambda}(^1S_0) &=& -1.555\ [fm]\ ,\ \ 
 r_{\Lambda\Lambda}(^1S_0)  =   3.617 [fm]\ .    
\label{le.3}\end{eqnarray*}
\begin{eqnarray*}
 a_{\Xi N}(^3S_1) &=& +122.5\ [fm]\ ,\ \ 
 r_{\Xi N}(^3S_1)  =   2.083\ [fm]\ .    
\label{le.4}\end{eqnarray*}
\begin{table}
\caption{ $I=0$: Inverse-scattering-length and effective-range matrices at the $\Xi N$  
         threshold. The order of the states (1-2) reads $\Lambda\Lambda(^1S_0),
         \Xi N(^1S_0)$. 
         The dimension of the matrix elements are in [fm]$^{-1}(A^{-1}$
         and [fm]$(R)$.}
\begin{ruledtabular}
\begin{tabular}{r|rr|rr}
  & \multicolumn{2}{c}{ ESC04a } & \multicolumn{2}{c}{ ESC04d }\\ 
  & \multicolumn{1}{c}{$A^{-1}$}&\multicolumn{1}{c}{$R$} 
  & \multicolumn{1}{c}{$A^{-1}$}&\multicolumn{1}{c}{$R$}\\
\colrule \colrule
 11  &--0.929  &  4.114  &--0.584  & 3.484 \\    
 12  &  0.667  &--0.318  &  0.840  & --1.666\\    
 22  &  1.108  &  1.494  &  0.470  & 1.626 \\    
\end{tabular}
\end{ruledtabular}
\label{tabscat-2a} 
\end{table}

\begin{table}
\caption{ $I=1$: Inverse-scattering-length and effective-range matrices at the $\Xi N$  
         threshold. The order of the states (1-2) reads $\Lambda\Lambda(^1S_0),
         \Xi N(^1S_0)$. 
         The dimension of the matrix elements are in [fm]$^{-1}(A^{-1}$
         and [fm]$(R)$.}
\begin{ruledtabular}
\begin{tabular}{r|rr|rr}
  & \multicolumn{2}{c}{ ESC04a } & \multicolumn{2}{c}{ ESC04d }\\ 
  & \multicolumn{1}{c}{$A^{-1}$}&\multicolumn{1}{c}{$R$} 
  & \multicolumn{1}{c}{$A^{-1}$}&\multicolumn{1}{c}{$R$}\\
\colrule \colrule
 11  &  0.281  &  2.752  & --0.121 & 4.878 \\    
 12  &--4.067  &--1.643  & --0.977 & 1.851 \\    
 22  &--1.603  &  2.888  & --0.380 & 2.028 \\    
\end{tabular}
\end{ruledtabular}
\label{tabscat-2b} 
\end{table}
\begin{table}
\caption{ $I=1$: Inverse-scattering-length and effective-range matrices at the $\Xi N$  
         threshold. The order of the states (1-2) reads $\Lambda\Lambda(^3S_1),
         \Lambda\Lambda(^3D_1),\Xi N(^3S_1)$. 
         The dimension of the matrix elements are in [fm]$^{-1-l-l'}(A^{-1}$
         and [fm]$^{1-l-l'}(R)$.}
\begin{ruledtabular}
\begin{tabular}{r|rr|rr}
  & \multicolumn{2}{c}{ ESC04a } & \multicolumn{2}{c}{ ESC04d }\\ 
  & \multicolumn{1}{c}{$A^{-1}$}&\multicolumn{1}{c}{$R$} 
  & \multicolumn{1}{c}{$A^{-1}$}&\multicolumn{1}{c}{$R$}\\
\colrule \colrule
 11  &  1.484  &  4.896  &   0.021 & 139.289 \\    
 12  &--0.856  &--11.999 &   0.015 & --517.222 \\    
 13  &  1.621  &--8.293  & --0.046 & --160.711 \\    
 22  &270.392  &--422.973 & --0.009 & 5933.937 \\    
 23  &  0.348  &  36.874 &   0.045 & 1471.645 \\    
 33  &--2.156  &  15.785 & --0.274 & 382.035 \\    
\end{tabular}
\end{ruledtabular}
\label{tabscat-2c} 
\end{table}
For $I=1$ we have for ESC04a:
\begin{eqnarray*}
 a_{\Xi N}(^1S_0) &=&  0.491\ [fm]\ ,\ \ 
 r_{\Xi N}(^1S_0)  =  -0.421\ [fm]\ .    
\label{le.5}\end{eqnarray*}
For $I=1$ we have for ESC04d:
\begin{eqnarray*}
 a_{\Xi N}(^1S_0) &=&  0.144\ [fm]\ ,\ \ 
 r_{\Xi N}(^1S_0)  =   4.670\ [fm]\ .    
\label{le.6}\end{eqnarray*}
For the $S=-2$ sector the results are given in Table~\ref{tabscat-2a}-\ref{tabscat-2c}.
The $\Lambda\Lambda(^1S_0)$ scattering lengths are found to be larger than 
in the NSC97 models, indicating an attractive $\Lambda\Lambda$ interaction,
which is strongest in ESC04a. 

The old experimental information seemed to indicate a
separation energy of $\Delta B_{\Lambda\Lambda}=4-5$ MeV,
corresponding to a rather strong attractive $\Lambda\Lambda$
interaction. As a matter of fact, an estimate for the $\Lambda\Lambda$
$^1S_0$ scattering length, based on such a value for $\Delta
B_{\Lambda\Lambda}$, gives $a_{\Lambda\Lambda}(^1S_0)\approx-2.0$
fm~\cite{Tan65,Bod65}.
However, in recent years the experimental information and interpretation 
of the ground state levels of 
$_{\Lambda\Lambda}^{\,\ 6}$He, $_{\Lambda\Lambda}^{\;10}$Be, and
$_{\Lambda\Lambda}^{\;13}$B~\cite{Dal89}, has been changed drastically.
This because of the Nagara-event \cite{Tak01}, 
identified uniquely as $_{\Lambda\Lambda}^{\,\ 6}$He
\cite{Tak01}, which established that the $\Lambda\Lambda$-interaction is weaker
($\Delta B_{\Lambda\Lambda} \approx 1$ MeV).

In NSC97 \cite{RSY99} we could only increase the attraction in
the $\Lambda\Lambda$ channel by modifying the scalar-exchange
potential. If the scalar mesons are viewed as being mainly 
$q\bar{q}$ states, one finds that the (attractive) scalar-exchange
part of the interaction in the various channels satisfies
\begin{equation}
   |V_{\Lambda\Lambda}| < |V_{\Lambda N}| < |V_{N\!N}|,
\label{le.7} \end{equation}
suggesting indeed a rather weak $\Lambda\Lambda$-potential.
The NSC97 fits to the $Y\!N$ scattering data~\cite{RSY99} give values
for the scalar-meson mixing angle which seem to point to almost ideal
mixing for the scalars as $q\bar{q}$ states, and we found that an increased
attraction in the $\Lambda\Lambda$ channel would give rise to
(experimentally unobserved) deeply bound states in the $\Lambda N$ channel.
On the other hand, in the ESC04 models we have in principle more possibilities 
because of the presence of meson-pair poteials.
As one sees from the values of the $a_{\Lambda\Lambda}(^1S_0)$ 
in the ESC04 models of this paper we can produce 
the apparently required attraction in the $\Lambda\Lambda$
interaction without giving rise to $\Lambda N$ bound states (see below).
Notice that also in ESC04 we have scalar mixings
close to ideal ones, akin to NSC97.
The large values for the triplet effective range $r_t$ in $\Xi N$
is a simple reflection of the fact that the $^3S_1$
phase shift at small laboratory momenta is very small and only very
slowly increases in magnitude. 


\subsection{Bound states in $\protect\bm{S}$ waves }
\begin{figure}[hbt]
  \resizebox{3.5cm}{!}
 {\includegraphics[200, 50][400,750]{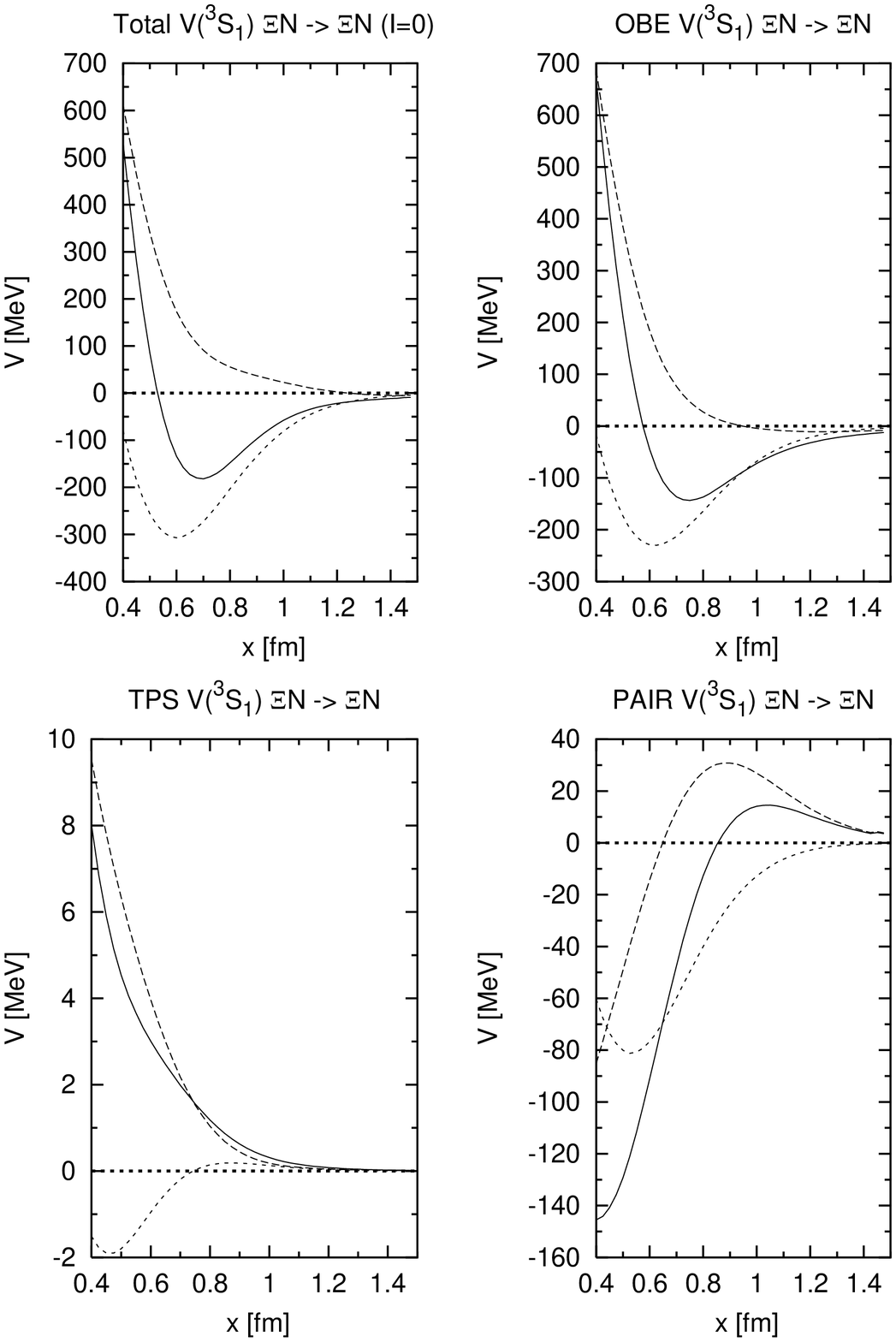}}
\caption{ESC04d: $\Xi N(^3S_1,I=0)$-potentials. The solid-, dashed-,
 and short-dashed-line are respectively the total-, central-, and 
 spin-spin-potentials.}
 \label{pot.esc041} 
 \end{figure}

A discussion of the possible bound-states, using the SU(3) content 
of the different $S=0, -1, -2$ channels is given in \cite{RSY99}.
As in \cite{RSY99}, for a general orientation, we list in Table~\ref{tabirrep} all
the irreps to which the various baryon-baryon channels belong.    
In contrast to the NSC97 models, we find almost no bound states in the 
ESC04a-d models. An exception is model ESC04d, where there occurs a
$\Xi N$ bound state in the $^3S_1$-$^3D_1$ coupled partial wave. 
From Table~\ref{tabirrep} one sees that this is a $\{8_a\}$-state, which
is a little bit surprising. This because the OBE-potential one expects
to be rather repulsive in the irrep $\{8_a\}$, see \cite{MRS89}.
However, in the ESC04 models this situation is changed because of the
contribution of the axial-vector-mesons, and the meson pairs, cfrm. Fig.~\ref{pot.esc041}.
The differences between ESC04a and ESC04d for this channel are in the 
OBE and MPE, and accidentally there is a bound-state in ESC04d but not
in ESC04a.

\begin{table}
\caption{SU(3) content of the different interaction channels.
         $S$ is the total strangeness and $I$ is the isospin.
         The upper half refers to the space-spin symmetric
         states $^3S_1$, $^1P_1$, $^3D$, \ldots, while the
         lower half refers to the space-spin antisymmetric
         states $^1S_0$, $^3P$, $^1D_2$, \ldots}
\begin{tabular}{cccc}
  \multicolumn{4}{c}{Space-spin symmetric}                              \\
 $S$ & $I$ & Channels                  & SU(3)-irreps                   \\
\colrule
   0 &   0 & $N\!N$                    & $\{10^*\}$                     \\
 --1 & 1/2 & $\Lambda N$, $\Sigma N$   & $\{10^*\}$, $\{8\}_a$          \\
     & 3/2 & $\Sigma N$                & $\{10\}$                       \\
 --2 &   0 & $\Xi N$                   & $\{8\}_a$                      \\
     &   1 & $\Xi N$, $\Sigma\Sigma$   & $\{10\}$, $\{10^*\}$, $\{8\}_a$\\
     &     & $\Sigma\Lambda$           & $\{10\}$, $\{10^*\}$           \\
  \multicolumn{4}{c}{Space-spin antisymmetric}                          \\
 $S$ & $I$ & Channels                  & SU(3)-irreps                   \\
\colrule
   0 &   1 & $N\!N$                    & $\{27\}$                       \\
 --1 & 1/2 & $\Lambda N$, $\Sigma N$   & $\{27\}$, $\{8\}_s$            \\
     & 3/2 & $\Sigma N$                & $\{27\}$                       \\
 --2 &   0 & $\Lambda\Lambda$, $\Xi N$, $\Sigma\Sigma$
                                       & $\{27\}$, $\{8\}_s$, $\{1\}$   \\
     &   1 & $\Xi N$, $\Sigma\Lambda$  & $\{27\}$, $\{8\}_s$            \\
     &   2 & $\Sigma\Sigma$            & $\{27\}$                       \\

\end{tabular}
\label{tabirrep}
\end{table}

From Table~\ref{tabirrep} one notices that this $\Xi N$-channel is not
a mixture of different SU(3)-irreps, and so form this point of view
simple. The same thing is for $S=-2$ only true for the 
$\Sigma\Sigma(^1S_0,I=2)$-channel. The other $S=-2$ channels are 
are mixtures of at least two irreps, which makes an analysis of the
presence or absence of bound states more difficult, as pointed out 
in \cite{RSY99}. For the models ESC04a-d we did not find any $S$-wave
bound states in these 'mixed' channels.


\subsection{Partial Wave Phase Parameters}    

For the {\it BB}-channels below the inelastic threshold we use for the parametrization
of the amplitudes the standard nuclear-bar phase shifts \cite{SYM57}.
The information on the
elastic amplitudes above thresholds is most conveniently given using the BKS-phases
\cite{Bry81,Kla83,Spr85}. For uncoupled partial waves, the elastic {\it BB}
$S$-matrix element is parametrized as
\begin{equation}
 S = \eta e^{2i \delta}\ ,\ \ \eta = \cos(2\rho)\ .
\label{pw.1} \end{equation}
For coupled partial waves the elastic {\it BB}-amplitudes are $2\times 2$-matrices.
The BKS $S$-matrix parametrization, which is of the type-S variety, is given by
\begin{equation}
 S = e^{i\delta} e^{i\epsilon} N\ e^{i\epsilon} e^{i\delta}\ ,
\label{pw.2} \end{equation}
where 
\begin{equation}
 \delta = \left(\begin{array}{cc} \delta_\alpha & 0 \\ 
          0 & \delta_\beta \end{array}\right)\ ,\ \
 \epsilon= \left(\begin{array}{cc} 0 & \epsilon \\ 
           \epsilon & 0 \end{array}\right)\ ,
\label{pw.3} \end{equation}
and $N$ is a real, symmetric matrix parametrize as
\begin{equation}
  N = \left(\begin{array}{cc} \eta_{11} & \eta_{12} \\ \eta_{12} & \eta_{22}
  \end{array}\right)\ .
\label{pw.4} \end{equation}
From the various parametrizations of the $N$-matrix, we choose the Kabir-Kermode
parametrization \cite{Kab87} to represent the $N$-matrix in the figures. 
Then, the $N$-matrix is given by the inelasticity 
parameters $(\alpha,\beta,\varphi)$, called $\rho$-parameters, as follows
\begin{equation}
  N = \left(\begin{array}{cc} \cos(2\alpha) & \sin(\varphi+\xi) \\              
                              \sin(\varphi+\xi) & \cos(2\beta)               
  \end{array}\right)\ , 
\label{pw.5} \end{equation}
where  
\begin{eqnarray}
&& \alpha = \pm \frac{1}{2}\cos^{-1}(\eta_{11})\ \ ,\ \ 
 \beta = \pm\frac{1}{2}\cos^{-1}(\eta_{22})\ , \nonumber\\
&& \varphi = \sin^{-1}(\eta_{12})-sgn(\eta_{12})\sin^{-1} Q\ \nonumber\\ 
&&   \xi     = sgn(\eta_{12})\sin^{-1} Q\ .        
\label{pw.6} \end{eqnarray}
Here 
\begin{equation}
 Q^2 = 1 - |\eta_{11}+\eta_{22}|+ \eta_{11} \eta_{22}\ .
\label{pw.7} \end{equation}

In Fig's~\ref{phasll0}-\ref{phassl1} the BKS-phases and coupling parameters
$(\alpha,\beta,\varphi)$ for ESC04d are shown.
In Fig~\ref{phasll0} and Fig.~\ref{phasxn1} we also show the $^1S_0$-phases (n.c.) 
for the case with no coupling to the other two-particle channels. 
For $\Lambda\Lambda$ the n.c.-curve 
shows that the potential is repulsive, which is mainly due to the $\{1\}$-irrep.
The attraction comes in particular from the coupling to the $\Xi N$-channel.

In the Tables~\ref{tab.bks0a}-\ref{tab.bks1h}, 
we give for the models ESC04a,d the inelasticity parameters $\rho$ and 
$\eta_{11}, \eta_{12}, \eta_{22}$, which
enable the reader to construct the $N$-matrix most directly.





\subsection{Total cross sections}
We next present the predictions for the total cross section for several
channels. 
We suppose always that the beam as well as the target are unpolarized.
Therefore, we incuded the statistical factors, which are $1/4$           
for the spin-singlet and $3/4$ for the spin-triplet case.


In Fig.~\ref{sigtll0} we present the elastic $\Lambda\Lambda$ and the 
inelastic $\Lambda\Lambda \rightarrow \Xi N$ total cross sections. 
Being dominantly S-wave, 
there is in principle has a (sharp) cusp at the $\Xi N$-threshold, 
i.e. $p_\Lambda=344.4$ MeV/$c^2$.

In Fig.~\ref{sigtxn1} we present the $\Xi N$ and $\Sigma\Lambda$ elastic and the, 
$\Xi N \rightarrow \Lambda\Lambda$ and $\Sigma\Lambda \rightarrow \Xi N$ 
inelastic total cross sections. 

For those cases where both baryons are charged, we do not
include the purely Coulomb contribution to the total cross section,
nor do we include the Coulomb interference to the nuclear amplitude.
The cross section is calculated by summing the contributions from
partial waves with orbital angular momentum up to and including $L=2$.
We find this to be sufficient for all the $S\neq0$ sectors; inclusion
of any higher partial waves has no significant effect. 
Inclusion of higher partial waves will shift the total cross section
to slightly higher values without changing the overall shape. Of course,
their inclusion would be necessary if a detailed comparison with real
experimental data were to be made.

 \begin{widetext}
 \begin{center}
 \begin{figure}[hbt]
 \resizebox{9.5cm}{14.25cm} 
 {\includegraphics[150,175][550,625]{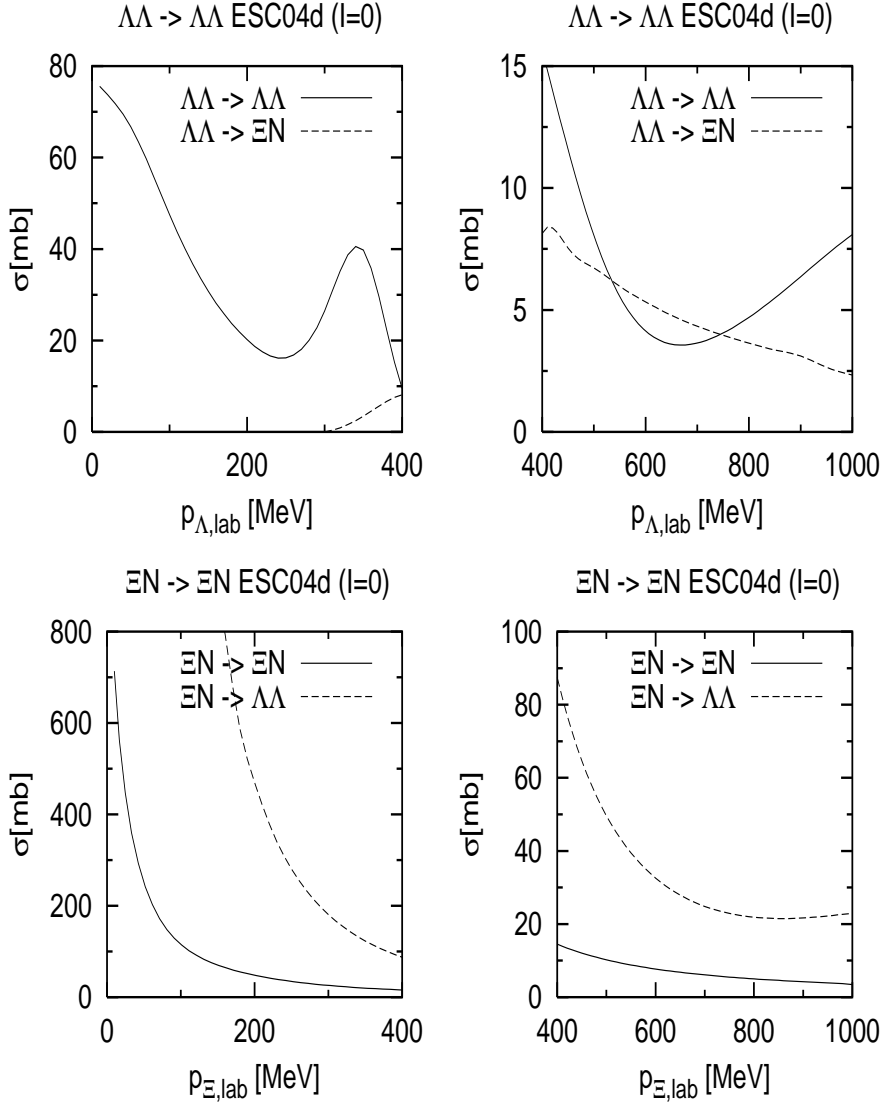}}
\caption{ESC04d $I=0$ $\sigma_T(\Lambda\Lambda$) and $\sigma_T(\Xi N)$.}  
 \label{sigtll0} 
 \end{figure}
 \end{center}
 \end{widetext}

 \begin{figure}   
 \resizebox{9.5cm}{14.25cm} 
 {\includegraphics[000,175][400,625]{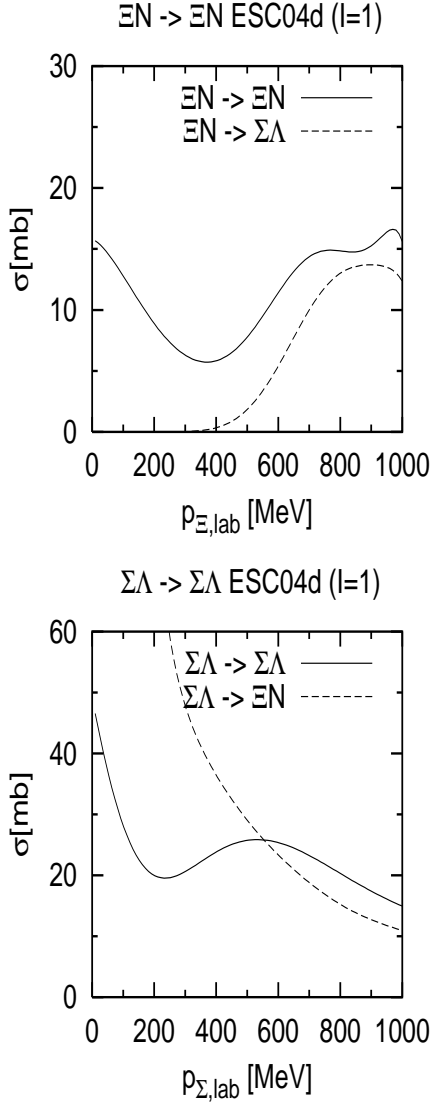}}
\caption{ESC04d $I=1$ $\sigma_T(\Xi N)$ and $\sigma_T(\Sigma\Lambda)$.}  
 \label{sigtxn1} 
 \end{figure}


In Table~\ref{tab.sigt0a} we show the $\Lambda\Lambda \rightarrow \Lambda\Lambda, \Xi N$
total X-sections as a function of the laboratory momentum $p_\Lambda$.
In Table~\ref{tab.sigt0b} we show the $\Xi N \rightarrow \Xi N, \Lambda\Lambda$
total X-sections as a function of the laboratory momentum $p_\Xi$.
In Table~\ref{tab.sigt1a} we show the $\Xi N \rightarrow \Xi N, \Sigma\Lambda$
total X-sections as a function of the laboratory momentum $p_\Xi$.
In Table~\ref{tab.sigt1b} we show the $I=1,L=0$ $\Sigma \Lambda \rightarrow 
\Sigma\Lambda, \Xi N, \Sigma\Sigma$
total X-sections as a function of the laboratory momentum $p_\Sigma$.


\begin{table}
\caption{$(I=0,L=0)$ Total X-sections $\Lambda\Lambda \rightarrow \Lambda\Lambda,
         \Xi N$ in [mb] as a function of the laboratory momentum $p_\Lambda$
         in [MeV]}
\begin{ruledtabular}
\begin{tabular}{r|rr|rr}
  & \multicolumn{2}{c}{ ESC04a } & \multicolumn{2}{c}{ ESC04d }\\ 
$p_{\Lambda}$  & \multicolumn{1}{c}{$\Lambda\Lambda$}&\multicolumn{1}{c}{$ \Xi N $} 
               & \multicolumn{1}{c}{$\Lambda\Lambda$}&\multicolumn{1}{c}{$ \Xi N $} \\
\colrule   
 10  & 447.92 & ---  & 75.59 & ---  \\
 50  & 326.93 & ---  & 66.71 & ---  \\
 100 & 171.59 & ---  & 47.39 & ---  \\
 200 &  50.31 & ---  & 18.51 & ---  \\
 300 &  17.72 & ---  &  7.45 & ---  \\
 350 &  11.30 & 0.72 &  6.87 & 2.56 \\
 400 &   5.90 & 1.42 &  2.74 & 4.37 \\
 500 &   1.50 & 1.19 &  0.95 & 3.44 \\
 600 &   0.22 & 0.89 &  1.04 & 2.57 \\
 700 &   0.07 & 0.67 &  1.48 & 1.99 \\
 800 &   0.34 & 0.51 &  1.87 & 1.60 \\
 900 &   0.73 & 0.39 &  2.13 & 1.33 \\
1000 &   1.10 & 0.30 &  2.34 & 1.02 \\
\end{tabular}
\end{ruledtabular}
\label{tab.sigt0a}
\end{table}
\begin{table}
\caption{$(I=0,L=0)$ Total X-sections $\Xi N \rightarrow \Xi N, \Lambda\Lambda$
         in [mb] as a function of the laboratory momentum $p_\Xi$
         in [MeV]}
\begin{ruledtabular}
\begin{tabular}{r|rr|rr}
  & \multicolumn{2}{c}{ ESC04a } & \multicolumn{2}{c}{ ESC04d }\\ 
$p_{\Xi}$ & \multicolumn{1}{c}{$\Xi N$}&\multicolumn{1}{c}{$\Lambda\Lambda$} 
          & \multicolumn{1}{c}{$\Xi N$}&\multicolumn{1}{c}{$\Lambda\Lambda$}\\
\colrule   
 10  & 184.19 & 278.99 & 708.56 &  187x10$^3$\\
 50  &  35.67 & 259.62 & 129.80 & 8496.85\\
 100 &  16.84 & 212.47 &  57.73 & 2073.20\\
 200 &   7.25 & 118.22 &  22.73 &  461.11\\
 300 &   4.08 &  63.66 &  12.12 &  168.70\\
 400 &   2.58 &  36.51 &   7.48 &   73.17\\
 500 &   1.76 &  22.90 &   5.07 &   34.51\\
 600 &   1.26 &  15.83 &   3.68 &   17.52\\
 700 &   0.94 &  12.02 &   2.81 &    9.98\\
 800 &   0.72 &   9.90 &   2.23 &    6.83\\
 900 &   0.56 &   8.71 &   1.83 &    5.76\\
1000 &   0.44 &   8.06 &   1.48 &    5.53\\
\end{tabular}
\end{ruledtabular}
\label{tab.sigt0b}
\end{table}
\begin{table}
\caption{$(I=1,L=0)$ Total X-sections $\Xi N \rightarrow \Xi N, \Sigma\Lambda$
         in [mb] as a function of the laboratory momentum $p_\Xi$
         in [MeV]}
\begin{ruledtabular}
\begin{tabular}{r|rr|rr}
  & \multicolumn{2}{c}{ ESC04a } & \multicolumn{2}{c}{ ESC04d }\\ 
$p_{\Xi}$ & \multicolumn{1}{c}{$\Xi N$}&\multicolumn{1}{c}{$\Sigma\Lambda$} 
          & \multicolumn{1}{c}{$\Xi N$}&\multicolumn{1}{c}{$\Sigma\Lambda$}\\
\colrule   
 10  & 297.54 & ----   &  15.68 & ---- \\    
 50  & 278.46 & ----   &  15.00 & ---- \\    
 100 & 233.23 & ----   &  13.11 & ---- \\    
 200 & 142.04 & ----   &   7.99 & ---- \\    
 300 &  87.03 & ----   &   3.91 & ---- \\    
 400 &  57.54 & ----   &   1.77 & ---- \\    
 500 &  41.08 & ----   &   0.85 & ---- \\    
 600 &  28.31 &   2.22 &   0.98 &  3.27\\
 700 &  20.64 &   3.63 &   2.55 &  3.21\\
 800 &  16.73 &   3.38 &   3.19 &  2.45\\
 900 &  14.10 &   3.27 &   3.62 &  2.20\\
 950 &  15.26 &   1.59 &  11.88 &  3.69\\
1000 &  12.60 &   2.34 &   8.45 &  1.85\\
\end{tabular}
\end{ruledtabular}
\label{tab.sigt1a}
\end{table}
\begin{table}
\caption{$(I=1,L=0)$ Total X-sections $\Sigma\Lambda \rightarrow \Sigma\Lambda, \Xi N,
         \Sigma\Sigma$
         in [mb] as a function of the laboratory momentum $p_\Sigma$
         in [MeV]}
\begin{ruledtabular}
\begin{tabular}{r|rrr|rrr}
  & \multicolumn{3}{c}{ ESC04a } & \multicolumn{3}{c}{ ESC04d }\\ 
$p_{\Sigma}$ & \multicolumn{1}{c}{$\Sigma\Lambda$}&\multicolumn{1}{c}{$\Xi N$} 
          & \multicolumn{1}{c}{$\Sigma\Sigma$}                              
          & \multicolumn{1}{c}{$\Sigma\Lambda$}&\multicolumn{1}{c}{$\Xi N$} 
          & \multicolumn{1}{c}{$\Sigma\Sigma$} \\                           
\colrule   
 10  & 671.52 &  31.36 & ---- & 1406.84 & 46.55 & ---- \\    
 50  & 127.24 &  29.40 & ---- &  223.90 & 36.19 &---- \\    
 100 &  58.38 &  26.51 & ---- &   84.93 & 24.49 & ---- \\    
 200 &  24.24 &  20.73 & ---- &   26.67 & 10.17 & ---- \\    
 300 &  13.63 &  16.18 & ---- &   12.38 &  4.87 & ---- \\    
 400 &   8.92 &  13.06 & ---- &    6.96 &  4.26 & ---- \\    
 500 &   6.56 &  10.74 & ---- &    4.50 &  5.50 & ---- \\    
 600 &   5.86 &  12.39 & ---- &    4.41 &  7.13 & ---- \\
 650 &   3.70 &  13.53 &  2.04&    3.85 &  7.92 &  0.81\\
 700 &   3.60 &  11.00 &  2.31&    2.50 &  8.35 &  0.46\\
 800 &   3.40 &   9.48 &  1.82&    1.85 &  9.24 &  0.17\\
 900 &   3.24 &   8.59 &  1.67&    1.56 &  9.76 &  0.26\\
1000 &   3.03 &   7.92 &  1.67&    1.36 &  9.94 &  0.54\\
\end{tabular}
\end{ruledtabular}
\label{tab.sigt1b}
\end{table}

\section{$\Xi N$ G-matrix interaction}
\label{sec:gmat1}

As demonstrated in our previous works~\cite{RSY99}~\cite{RY05},
the G-matrix theory is very convenient to explore the
features of {\it YN} and {\it YY} interaction models in nuclear medium.
Table XXV in Ref.\cite{RY05} demonstrates the basic features 
of the $\Xi N$ G-matrix interactions derived from ESC04a,b,c,d.
It is important, here, that some versions (ESC04c,d) lead to
the attractive $\Xi$-nucleus potentials $U_\Xi$, 
predicting the existence of $\Xi$ hypernuclei
owing to their strong $^{13}S_1$ attractions.
(A two-body spin- and isospin-state 
is represented by $^{(2I+1)(2S+1)}L_J$.)
In the present, the most reliable information for $U_\Xi$ is
considered to be given by the BNL-E885 experiment~\cite{E885},
in which they measured the missing mass spectra for 
the $^{12}$C$(K^-, K^+)$X reaction.  Reasonable agreement 
between this data and theory is realized by assuming 
a $\Xi$-nucleus potential $U_\Xi (\rho)=-V_0 f(r)$ with well 
depth $V_0 \sim 14$ MeV within the Wood-Saxon prescription,
named here as WS14.

\begin{table}[hbt]
\caption{$\Xi$ single particle energies $U_\Xi$ and conversion
widths $\Gamma_\Xi$ at normal density calculated with
ESC04d and NHC-D. $S$-state contributions in spin- and isospin-
states and total $P$-state contributions are also given.
All entries are in MeV.}
\begin{ruledtabular}
\begin{tabular}{l|rrrrr|rr}
  & $^{11}S_0$ & $^{13}S_1$ & $^{31}S_0$ & $^{33}S_1$ 
  & $P$ & $U_\Xi$ & $\Gamma_\Xi$ \\
\colrule
ESC04d($\alpha=0$)& 6.4 & $-$19.6 & 6.4 & $-$5.0 & $-$6.9 & $-$18.7 & 11.4 \\
ESC04d($\alpha=.18$)& 6.3 & $-$18.4 & 7.2 & $-$1.7 & $-$5.6 & $-$12.1 & 12.7 \\
NHC-D    & $-$2.6 & 0.7 & $-$2.3 & $-$0.4 & $-$16.8 & $-$21.4 & 1.1 \\
\end{tabular}
\end{ruledtabular}
\label{Gmat0}
\end{table}

Among the four versions of ESC04 models, only ESC04d seems
to be compatible with WS14. 
In Table~\ref{Gmat0}, we recapitulate the G-matrix result
for ESC04d, where the partial-wave contributions to $U_\Xi$
are shown. 
The most important is here that the attractive values of
$U_\Xi$ for ESC04d are due to the strong attractions
in the $^{13}S_1$ state.
The difference between the two versions of ESC04d specified
by values of $\alpha$ is as follows:
In Ref.\cite{RY05}, the medium-induced repulsion was
taken into account by changing masses of vector mesons
in medium with use of the parameter $\alpha_V$.
Then, it was shown that this effect plays important roles 
to reproduce nuclear saturation and $\Lambda$ well depth.
It is quite reasonable to take this effect into account
also in $\Xi$ hypernuclear systems:
A criterion for $\alpha_V$ is the value of 0.18
used successfully in $S=0$ and $-1$ cases.
However, we should not stick to this value, because there 
is no definite information for $\Xi$ hypernuclei experimentally 
in the present. A reasonable way for us is to consider 
it as a changeable parameter to study features of $\Xi$ states.
Hereafter, the parameter $\alpha_V$ is denoted as $\alpha$ simply.

In this work, the imaginary parts of G-matrices 
in $^{11}S_1$ and $^{13}P_J$ states are taken into account 
together with their real parts.
The imaginary parts are due to the energy-conserving transitions
from $\Xi N$ to $\Lambda \Lambda$ channels in nuclear medium.
For simplicity, we evaluate them in perturbation, 
the real parts being the same as those in Ref.\cite{RY05}.
Then, the conversion width $\Gamma_\Xi$ is obtained from
the imaginary part of $U_\Xi$ by multiplying $-2$.
The calculated values of $\Gamma_\Xi$ are also given 
in Table~\ref{Gmat0}.


It should be stressed here that the features of the $\Xi N$ 
interaction in ESC04d are distinctly different from 
those in OBE models. Among the Nijmegen OBE models, 
only the Nijmegen hard-core model D (NHC-D)~\cite{NRS77} 
is known to give the attractive value of $U_\Xi$ adequately
owing to its peculiar modeling, where octet scalar mesons 
are not taken into account.
For comparison, the result for NHC-D is also given
in Table~\ref{Gmat0}, where the hard-core radii $r_c$ are
taken so as to reproduce $\Xi$-nucleus interactions compatibly
to WS14: The value of $r_c$ in $^{11}S_0$ state is taken as 
0.53 fm, and those in the other channels as 0.47 fm.
The former value is chosen so that the $\Lambda \Lambda$
interaction in this channel is consistent with the
data of $^{\ 6}_{\Lambda \Lambda}$He.
The features of NHC-D are found to be quite different 
from those of ESC04d($\alpha$): The attractive value of
$U_\Xi$ in this case is dominated by the $P$-state 
contribution, and the calculated value of $\Gamma_\Xi$ 
is far smaller than those of ESC04d($\alpha$).

\begin{table}[hbt]
\caption{Parameters $c_{ij}$ in Eq.\ref{eq:YNG} for 
ESC04d($\alpha$).}
\begin{ruledtabular}
\begin{tabular}{cc|rrr|rrr}
\multicolumn{8}{l}{\hspace{4cm} Real Parts} \\
\colrule
&& \multicolumn{3}{c|}{$I=0$} & \multicolumn{3}{c|}{$I=1$} \\
\colrule
         & $c_{1j}^{(+)}$ & $-$896.7  & $-$181.5 & $-$1865. 
                 &    11.10  &    34.07 & $-$50.69 \\
$^{1}E$  & $c_{2j}^{(+)}$ &    1632.  &    356.4 &    3236.   
                 &    61.19  & $-$95.18 &    115.8 \\
         & $c_{3j}^{(+)}$ & $-$573.1  & $-$153.3 & $-$1457.   
                 &    3.896  &    83.62 & $-$63.83 \\
\colrule
         & $c_{1j}^{(+)}$ & $-$641.7  &    18.39 &    86.11   
                 & $-$35.84  &    33.02 &    .8333 \\
$^{3}E$  & $c_{2j}^{(+)}$ &    582.3  & $-$39.56 & $-$144.4   
                 &    20.39  & $-$96.59 & $-$5.139 \\
         & $c_{3j}^{(+)}$ & $-$197.9  &    64.61 &    63.89   
                 &    8.864  &    95.87 &    4.944 \\
\colrule
         & $c_{1j}^{(-)}$ &    312.4  & $-$13.64 &    26.39   
                 & $-$171.4  &    23.08 &    29.17 \\
$^{1}O$  & $c_{2j}^{(-)}$ &    75.00  &    13.86 & $-$19.58   
                 &    124.9  & $-$79.25 & $-$87.50 \\
         & $c_{3j}^{(-)}$ &    20.00  &    15.24 &    3.611   
                 & $-$21.81  &    111.7 &    55.55 \\
\colrule
         & $c_{1j}^{(-)}$ & $-$331.4  &    4.944 &    580.6   
                 & $-$108.2  &    39.69 &    43.06 \\
$^{3}O$  & $c_{2j}^{(-)}$ &    39.75  & $-$47.36 & $-$1326.   
                 &    24.25  & $-$120.1 & $-$113.6 \\
         & $c_{3j}^{(-)}$ &    81.25  &    98.96 &    760.4   
                 &    3.636  &    127.0 &    69.78 \\
\colrule
\end{tabular}
\end{ruledtabular}
\vskip 0.3cm
\begin{ruledtabular}
\begin{tabular}{cc|rrr}
\multicolumn{5}{l}{Imaginary Parts ($I=0$)} \\
\colrule
         & $c_{1j}^{(+)}$ & $-$292.0  & $-$916.9 &    569.4 \\
$^{1}E$  & $c_{2j}^{(+)}$ &    32.47  &    2096. & $-$1621. \\
         & $c_{3j}^{(+)}$ &    19.74  & $-$1152. &    1059. \\
\colrule
         & $c_{1j}^{(-)}$ & $-$10.55  &    1.619 & $-$523.2 \\
$^{3}O$  & $c_{2j}^{(-)}$ &    3.295  &    3.817 &    1178. \\
         & $c_{3j}^{(-)}$ & $-$1.242  & $-$8.286 & $-$655.1 \\
\end{tabular}
\end{ruledtabular}
\label{Gmat1}
\end{table}
Now, our concern is to investigate how appear the features of
ESC04d in level structures of various $\Xi$ hypernuclei.
For such an aim, it is convenient to represent $\Xi N$ G-matrix 
interactions in nuclear matter as density-dependent local 
potentials~\cite{Yam94}. Because sophisticated constructions 
of coordinate-space G-matrices are not necessary 
under our poor knowledge on $\Xi$ hypernuclei, 
we adopt here the following simple method: 
Our G-matrix interaction in each isospin- and spin-state
is given in a two-range Gaussian form
\begin{eqnarray}
&& G^{(\pm)}_{IS}(r,k_F) = (C_1^{(\pm)}+C_2^{(\pm)} k_F
+ C_3^{(\pm)} k_F^2)\cdot \nonumber\\ && \times \exp {(-(r/0.8)^2)}
+ C_0 \,\exp {(-(r/1.6)^2)} \ ,
\nonumber \\ 
&& C_i^{(\pm)} = \sum^3_{j=1} c_{ij}^{(\pm)}
\,\alpha^{j-1} \qquad {\rm for} \ \ i=1,2,3
             \label{eq:YNG}
\end{eqnarray}
where the density-dependence is represented as a function of 
a Fermi momentum $k_F$. The suffices $(+)$ and $(-)$ specify
even and odd states, respectively.
Parameters $C_i^{(\pm)}$ are determined as follows:
First, the outer-range part is fixed so as to simulate 
the tail part of the bare interaction.
The adopted values of $C_0$ are  
$-3.73$ MeV ($^{11}E$, $^{11}O$), 
$-5.29$ MeV ($^{13}E$, $^{13}O$),
$-3.34$ MeV ($^{31}E$, $^{31}O$)
and $-4.93$ MeV ($^{33}E$, $^{33}O$).
Next, the strengths of the inner-range parts are determined
so as to reproduce the partial-wave contributions to $U_\Xi$
in $(ISL)$ states. 
The $k_F$ dependence is determined by using the results
for the three values of $k_F=$ 1.35, 1.00, 0.80 fm$^{-1}$.
For convenience, the coefficients $C_i^{(\pm)} (i=1,2,3)$ 
in each $(IS)$ state are given as a quadratic function of 
$\alpha$, which represents all together the G-matrix 
interactions for ESC04d with various values of $\alpha$.
The parameters  $c_{ij}^{(\pm)}$ for ESC04d($\alpha$) are 
listed in Table~\ref{Gmat1}.

\section{Applications to $\Xi$ hypernuclei}
\label{sec:gmat2}

\subsection{A folding-model}

$\Xi$-nucleus potentials in finite systems are constructed by folding
our G-matrix interactions $G^{(\pm)}_{IS}(r,k_F)$ into nuclear-core 
density distributions with a local density approximation (LDA).
Taking into account the Lane term,
our potential $U_\Xi$ is given by

\begin{eqnarray}
&& U_\Xi({\bf r},{\bf r'})=
U_0({\bf r},{\bf r'})+U_1({\bf r},{\bf r'})
\,{\bf t_\Xi}\cdot {\bf T_c}/A_c\ ,
\nonumber \\
&& U_i({\bf r},{\bf r'}) = \delta({\bf r}-{\bf r'})
\int d{\bf r''} \rho({\bf r''})\,
 \left[V_i^{(+)}(|{\bf r}-{\bf r''}|;k_F)+ \right.\nonumber\\
  && \left. V_i^{(-)}(|{\bf r}-{\bf r''}|;k_F)\right]/2
 +  \rho({\bf r},{\bf r'})\left[ V_i^{(+)}(|{\bf r}-{\bf r'}|;k_F) -
 \right.\nonumber\\ && \left.
 V_i^{(-)}(|{\bf r}-{\bf r'}|;k_F)\right]/2
\ , \quad (i=0,1)\ ,
\end{eqnarray}

with $k_F({\bf r},{\bf r'})
=[3\pi^2/2\cdot(\rho({\bf r})+\rho({\bf r'}))/2]^{1/3}$.
The terms $V_i^{(\pm)}$ are expressed by combinations of 
$G^{(\pm)}_{IS}$ as
\begin{eqnarray}
V_0^{(\pm)}&=&(3G^{(\pm)}_{10}+3G^{(\pm)}_{01}
              +G^{(\pm)}_{00}+9G^{(\pm)}_{11})/16,
\nonumber \\
V_1^{(\pm)}&=&(G^{(\pm)}_{10}-3G^{(\pm)}_{01}
              -G^{(\pm)}_{00}+3G^{(\pm)}_{11})/4,
\end{eqnarray}
\noindent
where ${\bf T_c}$ and ${\bf t_\Xi}$ are isospins 
of a core nucleus and a $\Xi$ particle, respectively,
and $A_c$ is a mass number of a core nucleus.
Spin-dependent parts of $\Xi-$nucleus potentials
are not considered in our present studies.

Here, we study only the diagonal part of the 
${\bf t_\Xi}\cdot {\bf T_c}$ term, though the 
$\Xi^--\Xi^0$ mixing effect induced by
its non-diagonal part is important to specify
some feature of an underlying $\Xi N$ 
interaction~\cite{Lanskoy}.
Nuclear cores are assumed to be spherically symmetric, 
and density $\rho(r)$ and mixed density $\rho(r,r')$ are 
constructed from nuclear wave functions given by the 
density-dependent Hartree-Fock (DDHF) calculations with 
the Skyrme-III interaction~\cite{Skyrme3}. 
It should be noted here that the $\Xi N$ space-exchange 
parts are treated accurately in our treatment. 
In the present modeling, the difference between $\Xi^--$ 
and $\Xi^0-$nucleus potentials comes from the Lane term,
when the Coulomb interactions are switched off.

First, let us show the results for simple systems composed
of spin- and isospin-saturated nuclear cores attached by 
a $\Xi^-$ particle;
$^{12}$C+$\Xi^-$, $^{16}$O+$\Xi^-$, $^{28}$Si+$\Xi^-$,
$^{40}$Ca+$\Xi^-$, $^{90}$Zr+$\Xi^-$, where Coulomb interactions
between $\Xi^-$ and nuclear cores are taken into account.
In these systems, there is no contribution from the Lane term
except the case of $^{90}$Zr core.
In the left and right sides of Fig.1, 
full circles connected by solid lines show the single particle
(s.p.) energies of $\Xi^-$-bound states calculated with 
G-matrix interactions derived from ESC04d($\alpha=.18$) 
and NHC-D, respectively, as a function of ${A_c}^{-2/3}$.
The ``error bars'' in the figure present the calculated values
of conversion widths $\Gamma_\Xi$, 
though they are not visible in the case of NHC-D. 
The conversion widths for ESC04d are found to be 
remarkably larger than those for NHC-D, because the 
$\Xi N$-$\Lambda \Lambda$ coupling interaction in the 
$^{11}S_0$ state in the former is far stronger than that 
in the latter. This feature can be seen in our result for 
the double-$\Lambda$ nucleus $^{\ 6}_{\Lambda \Lambda}$He:
As shown in Table XXIV of Ref.\cite{RY05}, ESC04d brings about
a large value of the $\Xi N$ admixture probability $P_{\Xi N}$ in 
$^{\ 6}_{\Lambda \Lambda}$He due to the strong 
$\Lambda \Lambda$-$\Xi N$ coupling interaction. If this 
coupling is switched off in this case, no reasonable
$\Lambda \Lambda$ bound state can be obtained.

The open circles connected by dotted lines give the $s$-state
energies of $\Xi^0$.
It is noted that the Coulomb contributions to $\Xi^-$ binding
energies are substantial in large mass-number systems.
Hereafter, when a $\Xi$ particle can be bound without 
an assist from a $\Xi$-nucleus Coulomb interaction,
we call it a $\Xi$-nuclear bound state.
In the figure, $p$-states in $^{12}$C and $^{16}$O,
$d$-states in $^{28}$Si and $^{40}$Ca,
$f$- and $g$-states in $^{90}$Zr are so-called
Coulomb-assisted states. Namely, these $\Xi$ states become 
unbound when Coulomb interactions are switched off, though 
their wave functions deviate substantially from pure 
Coulomb ones. On the other hand, the $f$-state in $^{90}$Zr
for NHC-D is a $\Xi$-nuclear bound state.

ESC04d and NHC-D give rise to similar values of
$\Xi^-$ s.p. energies in the $^{12}$C core.
In the large mass-number region, however, 
the $\Xi^-$ s.p. energies for NHC-D are far deeper than 
those for ESC04d($\alpha=.18$), the reason why is because the 
$\Xi$-nucleus interaction for NHC-D is dominated 
by contributions from odd-state attractions.
There is no space-exchange term in OBE parts, because 
strangeness $-2$ cannot be carried by a single boson.
This is the reason why the odd-state interactions
in NHC-D are so attractive.

In the case of $^{90}$Zr+$\Xi^-$, the contributions 
from the Lane terms are +1.3 and +0.2 MeV for 
ESC04d($\alpha=.18$) and NHC-D, respectively.
It should be noted that the lane term in ESC04d is 
far stronger than that in NHC-D. 

Next, we study more realistic $\Xi$ hypernuclei produced by 
$p (K^-,K^+) \Xi^-$ reactions on available nuclear targets.
In Table~\ref{Gmat2}, our results for ESC04d($\alpha=.18$)
and ESC04d($\alpha=0$) are listed in some cases
of $N=Z$ targets ($^6$Li, $^{12}$C, $^{16}$O, $^{28}$Si,
$^{40}$Ca) and $N>Z$ targets ($^{11}$B, $^{27}$Al, $^{48}$Ca),
$Z$ and $N$ being proton and neutron numbers, respectively.
Here, we show the calculated values of $\Xi^-$ s.p.
energies $E_{\Xi^-}$, contributions $\Delta E_L$ and
$\Delta E_C$ from Lane terms and Coulomb interactions,
respectively, 
and conversion widths $\Gamma_{\Xi^-}$.
The obtained values of $\Xi^-$ $s$-state energies in 
$^{12}_{\Xi^-}$Be, being $-$4.1 and $-$5.5 MeV for 
ESC04d($\alpha=.18$) and ESC04d($\alpha=0$), respectively,
are comparable to the corresponding value $-$4.9 MeV for WS14.
The $S$-state interactions in ESC04d are rather attractive
in average owing to the strong attraction in the $^{13}S_1$
state. This feature is demonstrated by the fact that
there appear $\Xi$ hypernuclear states such as $^5_{\Xi^-}$H 
even in light $p$-shell systems.
In the other hand, there appears no $\Xi$-hypernuclear state
in the light $p$-shell region in the case of using NHC-D.

In the above cases, the Lane terms are in proportion to
$(N-Z+1)/4$, and work repulsively. Their contributions are 
found to be more repulsive in the cases of $N>Z$ targets. 
Especially, the $\Xi^-$ s.p. energies in $^{48}_{\Xi^-}$Ar
are noted to be shallower than those in $^{40}_{\Xi^-}$Ar 
because of the large repulsive contributions of the Lane term.
On the other hand, the Lane term derived from NHC-D
is far smaller than that from ESC04d.
In the largest case of the $\Xi^-$ state in $^{48}_{\Xi^-}$Ar,
for instance, we obtain the value of $\Delta E_L=0.27$ MeV
for NHC-D, which should be compared to the values 2.03 MeV 
($\alpha=.18$) and 2.16 MeV ($\alpha=0$) ESC04d
in Table~\ref{Gmat2}.
The strong Lane term in ESC04d is understood from 
the strong isospin dependence of the partial wave contribution,
as seen in Table~\ref{Gmat0}.

If $(K^-,K^0)$ reactions are realized in future,
we can expect to observe peculiar $\Xi$ hypernuclear states
predicted by ESC04d. As an example, the result for
$^{12}_{\Xi^-}$B is given in the bottom of 
Table~\ref{Gmat2}, which can be produced by the
$n(K^-,K^0)\Xi^-$ reaction on $^{12}$C target.
Here, the Lane terms are found to work attractively.
It is interesting that there appears the large difference
between $\Xi^-$ s.p. energies of $^{12}_{\Xi^-}$Be and
$^{12}_{\Xi^-}$B produced by $(K^-,K^+)$ and $(K^-,K^0)$
reactions on $^{12}$C target, respectively.

Thus, the $\Xi$ hypernuclear states produced by ESC04d
turn out to be of peculiar features: The attractive 
$\Xi$-nucleus interactions are realized by the strong 
$\Xi N$ attraction in the $^{13}S_1$ state, which brings 
about the strong Lane terms and produces $\Xi$-nuclear 
bound states in $s$- and light $p$-shell regions.
The strong $\Xi N$-$\Lambda \Lambda$ coupling interaction,
which is responsible to a reasonable $\Lambda \Lambda$
attraction, leads to rather large values of $\Gamma_\Xi$.

\subsection{A four-body $\Xi^0-\Xi^-$ mixed state}

The strong $^{13}S_1$ attraction in ESC04d makes it 
possible that there appear peculiar bound $\Xi$ states 
in a few body systems. 
%
As an example, let us study the features of the $\Xi$ four-body
system on the basis of ESC04d, which is observable in principle 
through $^4$He$(K^-,K^0)$ reactions.
We adopt here the coupled-channel model in the charge space,
which was formulated for the $^4_\Sigma$He system~\cite{Yamada92}.
In our present case, the mixing is taken into account
between $[\Xi^0+^3$H$]$ and $[\Xi^-+^3$He$]$ channels.
The basic coupled-channel equation given by Eq.(3.5) in
Ref.\cite{Yamada92} is solved variationally in the Gaussian base.
The folding potential between $\Xi$ and $3N$ cluster
($^3$H or $^3$He) is given as
\begin{eqnarray}
&7 \hspace{-5mm} U_{\Xi-3N}(R)=U_0(R) + U_\tau(R) ({\bf t_\Xi}\cdot{\bf t_{3N}})
+ \nonumber\\ &7 \hspace{-5mm} 
U_\sigma(R) ({\bf s_\Xi}\cdot{\bf s_{3N}})+ U_{\tau\sigma}(R)
({\bf t_\Xi}\cdot{\bf t_{3N}})({\bf s_\Xi}\cdot{\bf s_{3N}}) 
\end{eqnarray}
which is derived from our $\Xi N$ G-matrix interaction        
under the LDA.  Here, it should be noted that the spin 
dependence is taken into account exactly.
For the $3N$ core part we use the theoretical density distribution
$\rho(r)$ obtained from the three-body calculation~\cite{Ishikawa}. 
The space-exchange terms are taken into account by using 
some approximated expression for the mixed density 
$\rho(r,r')$~\cite{Ehime01}.

\newpage
\begin{widetext}
\begin{center}
\begin{figure}[hbt]
 \includegraphics*[width=14cm]{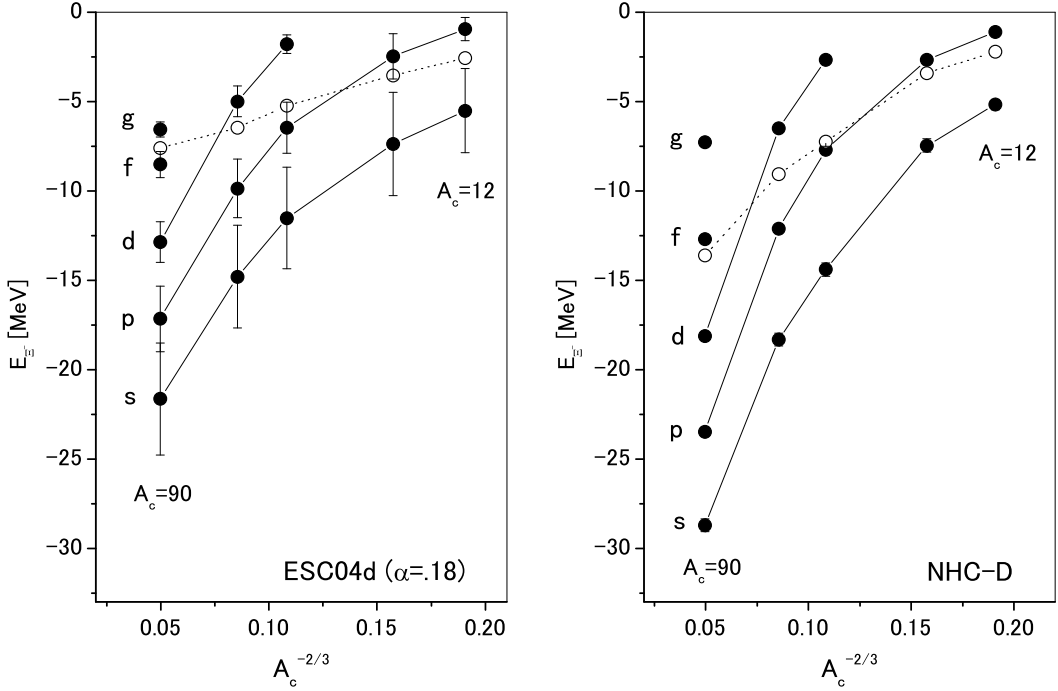}
\caption{$\Xi^-$ single particle energies for $^{12}$C+$\Xi^-$, 
$^{16}$O+$\Xi^-$, $^{28}$Si+$\Xi^-$, $^{40}$Ca+$\Xi^-$ and 
$^{90}$Zr+$\Xi^-$ are shown by full circles in the cases 
of ESC04d($\alpha=.18$) (left side) and NHC-D (right side).
Horizontal axes are for ${A_c}^{-2/3}$,
$A_c$ being a mass number of a core nucleus.
The ``error bars'' show calculated values of conversion widths.
Open circles show $\Xi^0$ $s$-state energies. }
\end{figure}

\begin{table}[hbt]
\caption{Calculated values of $\Xi^-$ single particle
energies $E_{\Xi^-}$ and conversion widths $\Gamma_\Xi$.
$\Delta E_L$ and $\Delta E_C$ are contributions from 
Lane terms and Coulomb interactions, respectively.
All entries are in MeV.}
\begin{ruledtabular}
\begin{tabular}{|ccc|rrrc|rrrc|}\hline
  &&& \multicolumn{4}{c|}{ESC04d($\alpha=.18$)} 
  & \multicolumn{4}{c|}{ESC04d($\alpha=0$)} \\
 \multicolumn{3}{|l|}{Target} & $E_{\Xi^-}$ & $\Delta E_L$ & $\Delta E_C$ 
  & $\Gamma_{\Xi^-}$ 
  & $E_{\Xi^-}$ & $\Delta E_L$ & $\Delta E_C$ & $\Gamma_{\Xi^-}$
  \\
  \colrule
$^{6}$Li& [$^5_{\Xi^-}$H]
             & $s$ & $-$0.6 & 0.45 & $-$0.57 & 1.4 
             & $-$0.8 & 0.56 & $-$0.69 & 1.8 \\    
$^{12}$C& [$^{12}_{\Xi^-}$Be]
             & $s$ & $-$4.1 & 0.51 & $-$2.35 & 4.3 
             & $-$5.5 & 0.62 & $-$2.58 & 5.2 \\    
$^{16}$O& [$^{16}_{\Xi^-}$C]
             & $s$ & $-$6.1 & 0.46 & $-$3.30 & 5.5 
             & $-$8.0 & 0.54 & $-$3.60 & 6.4 \\    
$^{28}$Si& [$^{28}_{\Xi^-}$Mg]
             & $s$ & $-$10.6 & 0.34 & $-$5.83 & 5.6 
             & $-$13.8 & 0.38 & $-$6.28 & 6.3 \\    
            && $p$ & $-$5.7 & 0.22 & $-$4.74 & 2.7 
             & $-$7.4 & 0.26 & $-$5.20 & 3.4 \\    
$^{40}$Ca& [$^{40}_{\Xi^-}$Ar]
             & $s$ & $-$14.0 & 0.26 & $-$7.94 & 5.7 
             & $-$17.8 & 0.28 & $-$8.44 & 6.3 \\    
            && $p$ & $-$9.2 & 0.19 & $-$6.80 & 3.2 
             & $-$11.5 & 0.22 & $-$7.34 & 3.8 \\    
            && $d$ & $-$4.4 & 0.12 &         & 1.7 
             & $-$5.7 & 0.15 &         & 2.1 \\    
\colrule
$^{11}$B& [$^{11}_{\Xi^-}$Li]
             & $s$ & $-$2.7 & 1.01 & $-$1.75 & 3.6 
             & $-$3.7 & 1.23 & $-$1.96 & 4.5 \\    
$^{27}$Al& [$^{27}_{\Xi^-}$Na]
             & $s$ & $-$9.6 & 0.70 & $-$5.37 & 5.4 
             & $-$12.7 & 0.78 & $-$5.83 & 6.2 \\    
            && $p$ & $-$4.9 & 0.44 & $-$4.29 & 2.6 
             & $-$6.4 & 0.53 & $-$4.74 & 3.3 \\    
$^{48}$Ca& [$^{48}_{\Xi^-}$Ar]
             & $s$ & $-$12.7 & 2.03 & $-$7.64 & 5.3 
             & $-$16.7 & 2.16 & $-$8.21 & 5.9 \\    
            && $p$ & $-$8.4 & 1.50 & $-$6.46 & 2.8 
             & $-$10.9 & 1.74 & $-$7.09 & 3.5 \\    
            && $d$ & $-$4.2 & 0.99 &         & 1.4 
             & $-$5.6 & 1.24 &         & 1.9 \\    
\colrule
$^{12}$C& [$^{12}_{\Xi^-}$B]
        & $s$ & $-$5.7 & $-$0.55 & $-$2.97 & 5.0 
        & $-$7.2 & $-$0.65 & $-$3.21 & 5.7 \\    
\end{tabular}
\end{ruledtabular}
\label{Gmat2}
\end{table}
\end{center}

\begin{center}
\begin{table}[hbt]
\caption{Energies $E_\Xi$ and conversion widths $\Gamma_\Xi$
in the pure $T=0$ states and the $[\Xi^0+^3$H$]-[\Xi^-+^3$He$]$
mixed states are shown together with probabilities of $\Xi^0$ 
components and $T=0$ components. Results are given in both 
cases of $J^\pi=0^+$, $1^+$. The values in parentheses are
obtained without the correction to the LDA ($R(A_c)$=1).}
\begin{ruledtabular}
\begin{tabular}{l|cccc|cccc}
& \multicolumn{4}{c }{$J^\pi=0^+$} 
& \multicolumn{4}{c }{$J^\pi=1^+$} \\
& $E_\Xi$ & $\Gamma_\Xi$ & $P_{\Xi^0}$ & $P_{T=0}$ 
& $E_\Xi$ & $\Gamma_\Xi$ & $P_{\Xi^0}$ & $P_{T=0}$ \\
& (MeV) & (MeV) &  (\%) & (\%) 
& (MeV) & (MeV) &  (\%) & (\%) \\
\colrule
ESC04d($\alpha=0$) &&&& &&&& \\ 
pure $T=0$  & $-$4.7  & 0.0     & ---  & --- 
            & $-$2.4  & 6.7     & ---  & --- \\
            &($-$2.9) &(0.0)    & ---  & --- 
            &($-$1.0) &(4.1)    & ---  & --- \\
$\Xi^0$-$\Xi^-$ mixed & $-$2.9  & 0.1 & 70.5 & 94.2 
                      & $-$1.4  & 4.7 & 93.2 & 73.6 \\
                      &($-$1.3) &(0.2)&(77.6)&(88.5)
                      &($-$0.4) &(2.1)&(98.1)&(61.3)\\
\colrule
ESC04d($\alpha=.18$) &&&& &&&& \\ 
pure $T=0$  & $-$3.9  & 0.0     & --- & --- 
            & $-$1.6  & 5.5     & --- & --- \\
            &($-$2.0) &(0.0)    & --- & --- 
            &($-$0.3) &(2.5)    & --- & --- \\
$\Xi^0$-$\Xi^-$ mixed & $-$2.2 & 0.2 & 73.5 & 92.1   
                      & $-$0.7 & 3.4 & 95.1 & 69.0 \\
                      &($-$0.6)&(0.2)&(84.3)&(81.3)  
                      &($-$0.0)&(0.6)&(99.6)&(53.3)\\
\end{tabular}
\end{ruledtabular}
\label{Gmat3}
\end{table}
\end{center}
\end{widetext}


Because the LDA seems to be rather problematic in our 
four-body system, let us investigate its reliability 
in the cases of light $\Lambda$ hypernuclei.
The $\Lambda N$ G-matrix interaction is obtained from
ESC04d($\alpha=0.1$), and the $\Lambda$-nucleus 
potentials are derived by folding them into
phenomenological density distributions of core nuclei.
Then, the experimental value of $B_\Lambda(^{13}_\Lambda$C) 
can be reproduced well under the above LDA.
When the same treatments are applied to lighter $\Lambda$ 
hypernuclei such as $^5_\Lambda$He, the LDA turns out to 
underestimate the $B_\Lambda$ values. We find that the 
experimental values are reproduced well by introducing
the following correction factor $R(A_c)$ into the LDA:
The local values of $k_F$, included in our G-matrix
interactions,  are taken as
$k_F({\bf r},{\bf r'}) = R(A_c)\ 
[3\pi^2/2\cdot(\rho({\bf r})+\rho({\bf r'}))/2]^{1/3}$
with $R(A_c)= 1.0-0.016\,(12-A_c)$.
In the region of $A_c<12$, this correction makes the 
$\Lambda$-nucleus interactions more attractive than
those in the simple LDA case. For instance, the calculated 
values of $B_\Lambda(^{5}_\Lambda$He) are 3.0 MeV and 1.8 MeV,
respectively, with and without this correction factor.
Thus, it is reasonable to use the same correction factor 
$R(A_c)$ in our calculations for the $\Xi$ four-body system.

In Table~\ref{Gmat3}, our results are given in the cases of 
ESC04($\alpha=0$) and ESC04($\alpha=.18$), $T$ being
sum of isospins of $\Xi$ and $3N$ clusters.
Here, the conversion widths are calculated in perturbation.
The values in parentheses here are obtained without
introducing the correction factor $R(A_c=3)$:
Even in this case, our following conclusions are not 
changed qualitatively.
First, let us remark on the fact that $\Xi$-nuclear bound 
states are obtained clearly in the case of pure $T=0$ states, 
where the Coulomb interactions are not taken into account. 
The reason can be understood as follows:
The $\Xi-3N$ folding interaction in the $(T,J^\pi)=(0,0^+)$ 
state is related to the $\Xi N$ interaction 
$G_{IS}$ through $3\cdot (G_{01}+G_{10})/2$.
Then, the strong attraction $G_{01}$ in ESC04d gives rise
to the substantial $\Xi-3N$ attraction. 
The $\Xi-3N$ interaction in the $(0,1^+)$ state
given by $3\cdot (3G_{11}+2G_{01}+G_{00})/6$, 
is less attractive than that in the $0^+$ state
because of the smaller weight of $G_{01}$.
On the other hand, the conversion width $\Gamma_\Xi$ is
determined dominantly by the imaginary part of $G_{00}$.
The reason why the obtained value of $\Gamma_\Xi$ 
in $(0,0^+)$ state is very small is because the $G_{00}$
component has no contribution to the $\Xi-3N$ interaction
in the this state.

Next, let us solve the $[\Xi^0+^3$H$]\otimes[\Xi^-+^3$He$]$
coupled-channel problem in the charge-space,
where the mass difference between $[\Xi^0+^3$H$]$ and 
$[\Xi^-+^3$He$]$ states and the $\Xi^--^3$He Coulomb 
interaction are taken into account.
In the single-channel treatment for $\Xi^0+^3$H ($\Xi^-+^3$He),
we obtain no $\Xi^0$-bound state (only a $\Xi^-$ Coulomb-bound
state). When the $[\Xi^0+^3$H$]-[\Xi^-+^3$He] coupling
interaction is switched on, we obtain $\Xi$-nuclear 
bound state both in the $0^+$ and $1^+$ channels. As shown 
by the value of the $\Xi^0$ mixing probability $P_{\Xi^0}$, 
this state is dominated by the lower $\Xi^0+^3$H component.
Then, $\Xi^-$-dominated states are found in continuum.

It is confirmed that the above $T=0$ bound-state solutions
are reproduced by the coupled-channel equation in the
charge base, when the mass difference $\Delta$ is taken to be 
zero and the coulomb interaction is switched off. 
The probabilities of $T=0$ components in our coupled-channel
solutions are also given in the Table.
Our solutions turn out to be deviated considerably from 
isospin eigenstates. This situation is contrastive to
the fact that the bound $^4_\Sigma$He system is of almost 
pure $T=1/2$ component.
The reason is mainly because the mass difference 5.9 MeV 
between $[\Xi^0+^3$H$]$ and $[\Xi^-+^3$He$]$ states
is considerably larger than the mass difference 2.6 MeV
between $[\Sigma^++^3$H$]$ and $[\Sigma^0+^3$He$]$ states.


\section{SUMMARY AND CONCLUSION}
\label{sec:conc}
The ESC04 models potentials presented here are a major step in constructing
the baryon-baryon interactions for scattering and hypernuclei in the
context of broken SU(3)$_F$ symmetry using, apart from an extremely simple
gaussian repulsion from the Pomeron, only meson-exchange for the dynamics.
The potentials are based on
(i) One-boson-exchanges, where the coupling constants at the
baryon-baryon-meson vertices are restricted by the broken SU(3) symmetry,
(ii) Two-pseudoscalar exchanges, (iii) Meson-Pair exchanges.
Each type of meson exchange (pseudoscalar, vector, axial-vector, scalar)
contains five free parameters: a singlet coupling constant, an octet
coupling constant, the $F/(F+D)$ ratio $\alpha$, a meson-mixing angle,
and for ESC04a-b a $\Delta_{FSB}$-parameter, which describes an SU(3)-symmetry
breaking of the meson couplings.
The potentials are regularized with gaussian cut-off parameters,
which provide a few additional free parameters.

Although we performed truly simultaneous fits to the {\it N\!N} and {\it Y\!N} data, 
effectively
most of these parameters are determined in fitting the rich and    
accurate {\it N\!N} scattering data, while the remaining ones are fixed
by fitting also the (few) {\it Y\!N} scattering data. This still leaves
enough freedom to construct the different models, ESC04a through ESC04d.
The distinction being using the different options w.r.t. $\Delta_{FSB}$ and
$a_{PV}$ (see \cite{RY05} for the definitions). These options are: (i) 
$\Delta_{FSB} \neq 0$ (ESC04a,b), or $\Delta_{FSB}=0$ (ESC04c,d); and (ii)
$a_{PV}=0.5$ (ESC04a,c), or $a_{PV}=1.0$ (ESC04b,d).    
They all describe the {\it N\!N} and {\it Y\!N} data equally well, but differ
on a more detailed level. The assumption of SU(3) symmetry then
allows us to extend these models to the higher strangeness channels
(i.e., {\it YY} and all interactions involving cascades), without the
need to introduce additional free parameters. 
Like the NSC97 models, the ESC04 models are very powerful models of this kind,
and the very first realistic ones.

In order to illustrate the basic properties of these potentials,
we have presented results for scattering lengths, possible bound
states in $S$-waves, and total cross sections. 
Although the four versions ESCa,b,c,d reproduce the $NN$ and $YN$ data 
equally well, there appear considerable differences in hypernuclear 
structures, especially in $S=-2$ systems. 
A typical example can be seen in their $\Xi N$ sectors: The derived
$\Xi$-nucleus potentials are different from each other even 
qualitatively. Then, it is quite important that one of the solutions
(ESC04d) in the ESC modeling predicts the existence of $\Xi$-hypernuclei
consistently with the indication given by the BNL-E885 experiment.
The $\Xi$-nucleus attraction derived from ESC04d is owing to the 
situation that the $\Xi N$ interaction in the $^3S_1$ ($^{33}S_1$)
state is substantially attractive (not strongly repulsive). This
feature is intimately related to its strong Lane term.
The mass dependence of $\Xi$ hypernuclei predicted by ESC04d is rather 
different from that by the OBE model such as NHC-D. The most striking
is that the peculier $\Xi N$ hypernuclear states are obtained by
ESC04d even in $s$- and light $p$-shell regions.

We finally mention that these ESC04 potentials also provide an
excellent starting point for calculations on multi-strange systems.
For that purpose it is necessary that we extend this work to the
$S=-3,4$-systems, i.e. comprising all $\{8\}\otimes\{8\}$ baqryon-baryon
states.


 \begin{widetext}
\acknowledgments
We thank Professor T. Motoba for stimulating discussions. Th.A.R. would like to thank 
P.M.M.\ Maessen and V.G.J. Stoks, for their collaboration in constructing 
the soft-core $S=-2$ OBE-model.
Y.Y. is most grateful to Professor D.E. Lanskoy for his enlightening and very
constructive comments on $\Xi$-hypernuclei.

\appendix 

\section{Baryon-baryon channels and SU(3)-irreps}
\label{app:C}
In Table~\ref{tab.irrep1} and Table~\ref{tab.irrep2}
we give the relation between the potentials
on the isospin basis and the potentials in the SU(3)-irreps.

\begin{table}
\caption{ SU(3)-contents of the various potentials 
          on the isospin basis.}
\begin{ruledtabular}
\begin{tabular}{ccl} 
\multicolumn{3}{c}{Space-spin antisymmetric states $^{1}S_{0},
                    \ ^{3}P,\ ^{1}D_{2},...$} \\
\colrule
$ \Lambda \Lambda \rightarrow \Lambda\Lambda $  & $ I=0 $ & 
$V_{\Lambda\Lambda,\Lambda\Lambda}
 = \frac{1}{40}\left(27 V_{27}+8 V_{8_{s}}+5 V_{1}\right)$\\
$ \Lambda \Lambda \rightarrow \Xi N  $  & ,, & $V_{\Lambda\Lambda,\Xi N}
 = \frac{-1}{40}\left(18 V_{27}-8 V_{8_{s}}-10 V_{1}\right)$\\
$ \Lambda \Lambda \rightarrow \Sigma\Sigma $  & ,, & $V_{\Lambda\Lambda,\Sigma\Sigma}
 =\frac{\sqrt{3}}{40}\left(-3 V_{27}+8 V_{8_{s}}-5 V_{1}\right)$\\
$ \Xi N   \rightarrow \Xi N  $  & ,, & $V_{\Xi N ,\Xi N}
 = \frac{1}{40}\left( 12 V_{27}+8 V_{8_{s}}+20 V_{1}\right)$\\
$ \Xi N   \rightarrow \Sigma\Sigma $  & ,, & $V_{\Xi N ,\Sigma\Sigma}
 =\frac{\sqrt{3}}{40}\left( 2 V_{27}+8 V_{8_{s}}-10 V_{1}\right)$\\
$ \Sigma\Sigma \rightarrow \Sigma\Sigma  $  & ,, & $V_{\Sigma\Sigma,\Sigma\Sigma}
 = \frac{1}{40}\left( V_{27}+24 V_{8_{s}}+15 V_{1}\right)$\\
\colrule
$ \Xi N   \rightarrow \Xi N   $  & $I=1$ & $V_{\Xi N ,\Xi N }
 = \frac{1}{5}\left( 2 V_{27}+3 V_{8_{s}}\right)$\\
$ \Xi N   \rightarrow \Lambda\Sigma  $  & ,,    & $V_{\Xi N ,\Lambda\Sigma}
 = \frac{\sqrt{6}}{5}\left( V_{27} - V_{8_{s}}\right)$\\
$ \Lambda\Sigma  \rightarrow \Lambda\Sigma  $  & ,,    & $V_{\Lambda\Sigma,\Lambda\Sigma}
 = \frac{1}{5}\left( 3 V_{27}+2 V_{8_{s}}\right)$\\
\colrule
$ \Sigma\Sigma  \rightarrow \Sigma\Sigma  $  & $I=2$ & $V_{\Sigma\Sigma,\Sigma\Sigma}
 = V_{27}$\\
\end{tabular}
\end{ruledtabular}
\label{tab.irrep1}
\end{table}
 
\begin{table}
\caption{ SU(3)-contents of the various potentials\\
 on the isospin         basis. }   
\begin{ruledtabular}
\begin{tabular}{ccl} 
\multicolumn{3}{c}{Space-spin symmetric states $^{3}S_{1},\ ^{1}P_{1},
                   \ ^{3}D,...$} \\
\colrule
$ \Xi N   \rightarrow \Xi N  $  & $ I=1$ & $V_{\Xi N,\Xi N}
 = \frac{1}{3}\left(V_{10}+ V_{10^{*}} + V_{8_{a}}\right)$\\
$ \Xi N   \rightarrow \Lambda\Sigma $  & $ ,,  $ & $V_{\Xi N ,\Lambda\Sigma}
 = \frac{\sqrt{6}}{6}\left(V_{10}- V_{10^{*}} \right)$\\
$ \Xi N   \rightarrow \Sigma\Sigma $  & $ ,, $ & $V_{\Xi N,\Sigma\Sigma}
 = \frac{\sqrt{2}}{6}\left(V_{10}+ V_{10^{*}} -2 V_{8_{a}}\right)$\\
$ \Lambda\Sigma  \rightarrow \Lambda\Sigma $  & $ ,, $ & $V_{\Lambda\Sigma,\Lambda\Sigma}
 = \frac{1}{2}\left(V_{10}+ V_{10^{*}} \right)$\\
$ \Lambda\Sigma  \rightarrow \Sigma\Sigma $  & $ ,, $ & $V_{\Lambda\Sigma,\Sigma\Sigma}
 = \frac{\sqrt{3}}{6}\left(V_{10}- V_{10^{*}} \right)$\\
$ \Sigma\Sigma  \rightarrow \Sigma\Sigma $  & $ ,, $ & $V_{\Sigma\Sigma,\Sigma\Sigma}
 = \frac{1}{6}\left(V_{10}+ V_{10^{*}}+4 V_{8_{a}} \right)$\\
\colrule
$ \Xi N   \rightarrow \Xi N  $  & $ I=0$ & $V_{\Xi N,\Xi N}
 =  V_{8_{a}}$\\
\end{tabular}
\end{ruledtabular}
\label{tab.irrep2}
\end{table}

\section{Meson-pair coupling constants}
\label{app:A}
In Table~\ref{prcop04a} and Table~\ref{prcop04d} we give the MPE-couplings
for model ESC04a and ESC04d respectively.

\begin{table}
\caption{Pair coupling constants for model ESC04a, divided by
         $\protect\sqrt{4\pi}$. $I(M)$ refers to the isospin of the pair $M$ with 
         quantum-numbers $J^{PC}$.}
\begin{ruledtabular}
\begin{tabular}{ccccrrrrccrrrr}
 Pair & $J^{PC}$ & Type & $I(M)$ & 
 $N\!N\!M$ & $\Sigma\Sigma M$ & $\Sigma\Lambda M$ & $\Xi\Xi M$
 & $\hspace*{2ex}$
 & $I(M)$ & $\Lambda N\!M$ & $\Lambda\Xi M$ & $\Sigma N\!M$ & $\Sigma\Xi M$ \\
\colrule
$\pi\eta$& $0^{++}$& $g$ 
    & $ 1 $  & --0.1860  & --0.3720  &   0.0000  & --0.1860 
    && $1/2$ &   0.3222  & --0.3222  &   0.1860  &   0.1860  \\
  & & & $ 0$ & --0.3222  &   0.0000  &   0.0000  &   0.3222  
    &&    &   &   &   &  \\
$\pi\pi$&$1^{--}$& $g$ 
    & $1$    & --0.0024  & --0.0049  &   0.0000  & --0.0024 
    && $1/2$ &   0.0042  & --0.0042  &   0.0024  &   0.0024  \\
  &&& $ 0 $  & --0.0042  &   0.0000  &   0.0000  &   0.0042  
    &&    &   &   &   &  \\
$\pi\pi$&$1^{--}$& $f$ 
    & $1$    &   0.1310  &   0.1048  &   0.0908  & --0.0262 
    && $1/2$ & --0.1361  &   0.0454  &   0.0262  & --0.1310  \\
  &&& $ 0 $  &   0.0454  & --0.0908  &   0.0908  & --0.1361  
    &&    &   &   &   &  \\
$\pi\rho$&$1^{++}$& $g$ 
    & $1$    &   0.8864  &   1.1404  &   0.3651  &   0.2540 
    && $1/2$ & --1.1702  &   0.8051  & --0.2540  & --0.8864  \\
  &&& $ 0 $  &   0.8051  & --0.3651  &   0.3651  & --1.1702  
    &&    &   &   &   &  \\
$\pi\sigma$&$1^{++}$& $g$ 
    & $1$    & --0.0241  & --0.0310  & --0.0099  & --0.0069 
    && $1/2$ &   0.0318  & --0.0219  &   0.0069  &   0.0241  \\
  &&& $ 0 $  & --0.0219  &   0.0099  & --0.0099  &   0.0318  
    &&    &   &   &   &  \\
$\pi\omega$&$1^{+-}$& $g$ 
    & $1$    & --0.1722  & --0.1608  & --0.1060  &   0.0114 
    && $1/2$ &   0.1923  & --0.0862  & --0.0114  &   0.1722  \\
  &&& $ 0 $  & --0.0862  &   0.1060  & --0.1060  &   0.1923  
    &&    &   &   &   &  \\[2mm]
\end{tabular}
\end{ruledtabular}
\label{prcop04a}
\end{table}

\begin{table}
\caption{Pair coupling constants for model ESC04d, divided by
         $\protect\sqrt{4\pi}$. $I(M)$ refers to the isospin of the pair $M$ with 
         quantum-numbers $J^{PC}$.}
\begin{ruledtabular}
\begin{tabular}{ccccrrrrccrrrr}
 Pair & $J^{PC}$ & Type & $I(M)$ & 
 $N\!N\!M$ & $\Sigma\Sigma M$ & $\Sigma\Lambda M$ & $\Xi\Xi M$
 & $\hspace*{2ex}$
 & $I(M)$ & $\Lambda N\!M$ & $\Lambda\Xi M$ & $\Sigma N\!M$ & $\Sigma\Xi M$ \\
\colrule
$\pi\eta$& $0^{++}$& $g$ 
    & $ 1 $  & --0.0971  & --0.1942  &   0.0000  & --0.0971 
    && $1/2$ &   0.1682  & --0.1682  &   0.0971  &   0.0971  \\
  & & & $ 0$ & --0.1682  &   0.0000  &   0.0000  &   0.1682  
    &&    &   &   &   &  \\
$\pi\pi$&$1^{--}$& $g$ 
    & $1$    &   0.0303  &   0.0607  &   0.0000  &   0.0303  
    && $1/2$ & --0.0526  &   0.0526  & --0.0303  & --0.0303  \\
  &&& $ 0 $  &   0.0526  &   0.0000  &   0.0000  & --0.0526
    &&    &   &   &   &  \\
$\pi\pi$&$1^{--}$& $f$ 
    & $1$    &   0.1390  &   0.1112  &   0.0963  & --0.0278 
    && $1/2$ & --0.1444  &   0.0481  &   0.0278  & --0.1390  \\
  &&& $ 0 $  &   0.0481  & --0.0963  &   0.0963  & --0.1444  
    &&    &   &   &   &  \\
$\pi\rho$&$1^{++}$& $g$ 
    & $1$    &   0.8344  &   0.9567  &   0.4111  &   0.1224 
    && $1/2$ & --1.0341  &   0.6230  & --0.1224  & --0.8344  \\
  &&& $ 0 $  &   0.6230  & --0.4111  &   0.4111  & --1.0341  
    &&    &   &   &   &  \\
$\pi\sigma$&$1^{++}$& $g$ 
    & $1$    & --0.0411  & --0.0471  & --0.0202  & --0.0060 
    && $1/2$ &   0.0509  & --0.0307  &   0.0060  &   0.0411  \\
  &&& $ 0 $  & --0.0307  &   0.0202  & --0.0202  &   0.0509  
    &&    &   &   &   &  \\
$\pi\omega$&$1^{+-}$& $g$ 
    & $1$    & --0.1690  & --0.1685  & --0.0978  &   0.0005 
    && $1/2$ &   0.1948  & --0.0970  & --0.0005  &   0.1690  \\
  &&& $ 0 $  & --0.0970  &   0.0978  & --0.0978  &   0.1948  
    &&    &   &   &   &  \\[2mm]
\end{tabular}
\end{ruledtabular}
\label{prcop04d}
\end{table}

\section{BKS-phase parameters}              
\label{app:B}
In Tables~\ref{tab.bks0a}-\ref{tab.bks1h} 
we display the BKS-phase parameters for model ESC04a and ESC04d.
\begin{table}[hbt]
\caption{$^1S_0(\Lambda\Lambda \rightarrow \Lambda\Lambda)$ BKS-phase parameters
         in [degrees] as a function of the laboratory momentum $p_\Lambda$
         in [MeV]}
\begin{ruledtabular}
\begin{tabular}{r|rr|rr}
  & \multicolumn{2}{c|}{ ESC04a } & \multicolumn{2}{c}{ ESC04d }\\ 
$p_{\Lambda}$  & \multicolumn{1}{c}{$\delta(^1S_0)$}&\multicolumn{1}{c|}{$\rho(^1S_0)$} 
               & \multicolumn{1}{c}{$\delta(^1S_0)$}&\multicolumn{1}{c}{$\rho(^1S_0)$}\\
\colrule   
 10  &   5.49 & ---  &  2.25 & ---  \\
 50  &  24.12 & ---  & 10.64 & ---  \\
 100 &  36.27 & ---  & 18.11 & ---  \\
 200 &  39.70 & ---  & 22.80 & ---  \\
 300 &  34.47 & ---  & 21.53 & ---  \\
 350 &  31.45 & 7.71 & 24.64 & 15.02\\
 400 &  25.89 &12.58 & 16.73 & 24.05\\
 500 &  16.11 &14.40 &  2.47 & 27.48\\
 600 &   6.38 &14.81 &--10.52& 28.69\\
 700 & --2.97 &14.82 &--20.94& 29.42\\
 800 & --7.43 &14.63 &--27.66& 30.01\\
 900 &--19.45 &14.20 &--29.86& 30.80\\
1000 &--26.26 &14.99 &--27.15& 30.39\\
\end{tabular}
\end{ruledtabular}
\label{tab.bks0a}  
\end{table}


\begin{table}     
\caption{$^1S_0(\Xi N \rightarrow \Xi N, I=0$ BKS-phase parameters
         in [degrees] as a function of the laboratory momentum $p_\Lambda$
         in [MeV]}
\begin{ruledtabular}
\begin{tabular}{r|rr|rr}
  & \multicolumn{2}{c|}{ ESC04a } & \multicolumn{2}{c}{ ESC04d }\\ 
$p_{\Xi}$  & \multicolumn{1}{c}{$\delta(^1S_0)$}&\multicolumn{1}{c|}{$\rho(^1S_0)$} 
               & \multicolumn{1}{c}{$\delta(^1S_0)$}&\multicolumn{1}{c}{$\rho(^1S_0)$}\\
\colrule   
 10  & --0.86 & 2.93 & --0.86 & 5.78 \\
 50  & --4.31 & 6.49 & --4.32 &12.68 \\
 100 & --8.74 & 8.98 & --8.62 &17.42 \\
 200 &--17.84 &11.91 &--16.90 &22.83 \\
 300 &--26.52 &13.47 &--24.08 &25.67 \\
 350 &--30.23 &13.95 &--26.90 &26.57 \\
 400 &--33.09 &14.29 &--28.96 &27.25 \\
 500 &--34.57 &14.69 &--30.33 &28.21 \\
 600 &--30.11 &14.84 &--27.75 &28.69 \\
 700 &--22.54 &14.83 &--22.00 &29.39 \\
 800 &--13.98 &14.71 &--14.36 &29.84 \\
 900 & --5.30 &14.47 & --6.25 &30.30 \\
1000 &   3.02 &14.20 & --2.65 &30.39 \\
\end{tabular}
\end{ruledtabular}
\label{tab.bks0b}  
\end{table}

\begin{table}     
\caption{$^1S_0(\Xi N \rightarrow \Xi N, I=1)$ BKS-phase parameters
         in [degrees] as a function of the laboratory momentum $p_\Xi$
         in [MeV]}
\begin{ruledtabular}
\begin{tabular}{r|rr|rr}
  & \multicolumn{2}{c|}{ ESC04a } & \multicolumn{2}{c}{ ESC04d }\\ 
$p_{\Xi}$      & \multicolumn{1}{c}{$\delta(^1S_0)$}&\multicolumn{1}{c|}{$\rho(^1S_0)$} 
               & \multicolumn{1}{c}{$\delta(^1S_0)$}&\multicolumn{1}{c}{$\rho(^1S_0)$}\\
\colrule   
 10  & --3.65 & ---  & --0.17 & ---  \\
 50  &--17.88 & ---  & --0.87 & ---  \\
 100 &--34.00 & ---  & --1.80 & ---  \\
 200 &--31.27 & ---  & --3.91 & ---  \\
 300 &--14.37 & ---  & --6.34 & ---  \\
 400 & --2.23 & ---  & --8.60 & ---  \\
 500 &   6.48 & ---  & --9.23 & ---  \\
 600 &   8.45 &14.50 &   3.28 & 25.85\\
 700 &  19.43 &23.36 &--19.59 & 31.54\\
 800 &  26.84 &25.98 &--27.01 & 30.59\\
 900 &  30.22 &27.37 &--29.78 & 29.25\\
1000 &  29.77 &28.19 &--30.40 & 27.65\\
\end{tabular}
\end{ruledtabular}
\label{tab.bks1a}  
\end{table}

\begin{table}     
\caption{$^1S_0(\Sigma\Lambda\rightarrow \Sigma\Lambda, I=1)$ BKS-phase parameters
         in [degrees] as a function of the laboratory momentum $p_\Sigma$
         in [MeV]}
\begin{ruledtabular}
\begin{tabular}{r|rr|rr}
  & \multicolumn{2}{c|}{ ESC04a } & \multicolumn{2}{c}{ ESC04d }\\ 
$p_{\Sigma}$   & \multicolumn{1}{c}{$\delta(^1S_0)$}&\multicolumn{1}{c|}{$\rho(^1S_0)$} 
               & \multicolumn{1}{c}{$\delta(^1S_0)$}&\multicolumn{1}{c}{$\rho(^1S_0)$}\\
\colrule   
 10  & --1.14 & 4.85 &   0.30 & 9.46 \\
 50  & --5.74 &10.68 &   1.43 & 20.08\\
 100 &--11.69 &14.73 &   2.58 & 26.15\\
 200 &--22.84 &19.76 &   3.41 & 30.81\\
 300 &--29.83 &22.96 &   0.86 & 31.57\\
 400 &--30.97 &25.13 & --4.86 & 31.06\\
 500 &--27.57 &26.63 &--12.01 & 30.09\\
 600 &--21.67 &27.65 &--19.01 & 28.82\\
 700 &--14.72 &28.35 &--24.85 & 27.20\\
 800 & --7.58 &28.79 &--29.05 & 25.12\\
 900 & --0.76 &29.06 &--31.60 & 22.40\\
1000 &   5.37 &29.20 &--32.89 & 18.92\\
\end{tabular}
\end{ruledtabular}
\label{tab.bks1b}  
\end{table}
\begin{table}     
\caption{ESC04a $^3S_1-^3D_1(\Xi N \rightarrow \Xi N, I=1)$ BKS-phase parameters
         in [degrees] as a function of the laboratory momentum $p_\Xi$
         in [MeV]}
\begin{ruledtabular}
\begin{tabular}{r|rrr|rrr}
$p_{\Xi}$      & \multicolumn{1}{c}{$\delta(^3S_1)$}&\multicolumn{1}{c}{$\epsilon_1$} 
               & \multicolumn{1}{c|}{$\delta(^3D_1)$} 
               & \multicolumn{1}{c}{$\eta_{11}$}&\multicolumn{1}{c}{$\eta_{12}$} 
               & \multicolumn{1}{c}{$\eta_{22}$}\\
\colrule   
 10  & --0.43 & 0.00 &   0.00 & ---   & ---  &  --- \\
 50  & --2.14 & 0.01 &   0.00 & ---   & ---  &  --- \\
 100 & --4.30 & 0.03 &   0.00 & ---   & ---  &  --- \\
 200 & --8.70 & 0.15 & --0.03 & ---   & ---  &  --- \\
 300 &--13.18 & 0.26 & --0.09 & ---   & ---  &  --- \\
 400 &--17.65 & 0.29 & --0.20 & ---   & ---  &  --- \\
 500 &--21.92 & 0.18 & --0.36 & ---   & ---  &  --- \\
 600 &--24.93 &--0.13& --0.59 &  0.97 & 0.00 & 1.00 \\
 700 &--30.37 &--0.66& --0.87 &  0.93 & 0.02 & 1.00 \\
 800 &--34.95 &--1.24& --1.22 &  0.91 & 0.04 & 1.00 \\
 900 &--39.29 &--1.76& --1.72 &  0.89 & 0.06 & 0.98 \\
1000 &--42.88 &--2.58& --2.14 &  0.92 & 0.09 & 0.97 \\
\end{tabular}
\end{ruledtabular}
\label{tab.bks1c}  
\end{table}

\begin{table}     
\caption{ESC04d $^3S_1-^3D_1(\Xi N \rightarrow \Xi N, I=1)$ BKS-phase parameters
         in [degrees] as a function of the laboratory momentum $p_\Xi$
         in [MeV]}
\begin{ruledtabular}
\begin{tabular}{r|rrr|rrr}
$p_{\Xi}$      & \multicolumn{1}{c}{$\delta(^3S_1)$}&\multicolumn{1}{c}{$\epsilon_1$} 
               & \multicolumn{1}{c|}{$\delta(^3D_1)$} 
               & \multicolumn{1}{c}{$\eta_{11}$}&\multicolumn{1}{c}{$\eta_{12}$} 
               & \multicolumn{1}{c}{$\eta_{22}$}\\
\colrule   
 10  &   0.48 & 0.00 &   0.00 & ---   & ---  &  --- \\
 50  &   2.36 & 0.00 &   0.00 & ---   & ---  &  --- \\
 100 &   4.39 & 0.00 &   0.00 & ---   & ---  &  --- \\
 200 &   6.65 &--0.03&   0.02 & ---   & ---  &  --- \\
 300 &   6.37 &--0.14&   0.10 & ---   & ---  &  --- \\
 400 &   4.26 &--0.40&   0.32 & ---   & ---  &  --- \\
 500 &   1.27 &--0.88&   0.74 & ---   & ---  &  --- \\
 600 & --1.76 &--1.66&   1.40 &  1.00 & 0.06 & 1.00 \\
 700 & --4.26 &--2.58&   2.02 &  0.99 & 0.09 & 0.99 \\
 800 & --4.88 &--3.93&   2.86 &  0.98 & 0.17 & 0.98 \\
 900 &   2.81 &--6.27&   3.81 &  0.95 & 0.22 & 0.97 \\
 950 &  38.03 &--8.72&   3.57 &  0.78 & 0.29 & 0.95 \\
1000 &--34.79 &--2.95&   4.33 &  0.49 & 0.07 & 0.98 \\
\end{tabular}
\end{ruledtabular}
\label{tab.bks1d}  
\end{table}
\begin{table}     
\caption{ESC04a $^3S_1-^3D_1(\Sigma\Lambda\rightarrow \Sigma\Lambda, I=1)$ 
         BKS-phase parameters in [degrees] as a function of the laboratory 
         momentum $p_\Xi$ in [MeV]}
\begin{ruledtabular}
\begin{tabular}{r|rrr|rrr}
\colrule   
$p_{\Sigma}$   & \multicolumn{1}{c}{$\delta(^3S_1)$}&\multicolumn{1}{c}{$\epsilon_1$} 
               & \multicolumn{1}{c|}{$\delta(^3D_1)$} 
               & \multicolumn{1}{c}{$\eta_{11}$}&\multicolumn{1}{c}{$\eta_{12}$} 
               & \multicolumn{1}{c}{$\eta_{22}$}\\
\colrule   
 10  &   0.40 & 0.00 &   0.00 &  1.00 & 0.00 & 1.00 \\
 50  &   1.90 & 0.00 &   0.00 &  0.98 & 0.00 & 1.00 \\
 100 &   3.17 & 0.00 & --0.03 &  0.97 &--0.01 & 1.00 \\
 200 &   3.31 & 0.02 & --0.28 &  0.94 &--0.04 & 1.00 \\
 300 &   0.99 & 0.10 & --0.63 &  0.93 &--0.05 & 1.00 \\
 400 & --2.39 & 0.27 & --0.75 &  0.92 &--0.03 & 1.00 \\
 500 & --5.01 & 0.59 & --0.35 &  0.91 &  0.02 & 1.00 \\
 600 &   9.74 & 1.73 &   2.79 &  0.81 &  0.32 & 0.93 \\
 700 &--22.23 & 1.60 & --1.17 &  0.76 &--0.08 & 0.98 \\
 800 &--25.56 & 0.94 &   0.13 &  0.76 &  0.00 & 0.94 \\
 900 &--29.18 & 1.05 &   0.49 &  0.74 &  0.07 & 0.88 \\
1000 &--32.87 & 1.78 & --0.17 &  0.72 &  0.13 & 0.81 \\
\end{tabular}
\end{ruledtabular}
\label{tab.bks1e}  
\end{table}
\begin{table}     
\caption{ESC04d $^3S_1-^3D_1(\Sigma\Lambda\rightarrow \Sigma\Lambda, I=1)$ 
         BKS-phase parameters in [degrees] as a function of the laboratory 
         momentum $p_\Xi$ in [MeV]}
\begin{ruledtabular}
\begin{tabular}{r|rrr|rrr}
$p_{\Sigma}$   & \multicolumn{1}{c}{$\delta(^3S_1)$}&\multicolumn{1}{c}{$\epsilon_1$} 
               & \multicolumn{1}{c|}{$\delta(^3D_1)$} 
               & \multicolumn{1}{c}{$\eta_{11}$}&\multicolumn{1}{c}{$\eta_{12}$} 
               & \multicolumn{1}{c}{$\eta_{22}$}\\
\colrule   
 10  &   0.38 & 0.00 &   0.00 &  1.00 & 0.00 & 1.00 \\
 50  &   1.78 & 0.00 &   0.00 &  1.00 & 0.00 & 1.00 \\
 100 &   2.86 & 0.00 & --0.02 &  1.00 &--0.01& 1.00 \\
 200 &   2.16 & 0.00 & --0.22 &  0.99 &--0.04& 1.00 \\
 300 & --1.69 &--0.02& --0.42 &  0.99 &--0.05& 1.00 \\
 400 & --7.31 &--0.08& --0.28 &  0.99 &--0.05& 1.00 \\
 500 &--13.56 &--0.16&   0.51 &  0.99 &--0.03& 0.99 \\
 600 &--19.45 & 0.08&   2.96 &  1.00 &  0.02& 0.93 \\
 650 &--22.61 & 1.47& --3.08 &  0.98 &--0.02& 0.90 \\
 700 &--25.59 & 0.86& --1.29 &  0.98 &--0.01& 0.96 \\
 800 &--31.15 & 0.47&   0.43 &  0.97 & 0.03 & 0.98 \\
 900 &--36.15 & 0.32&   1.67 &  0.95 & 0.07 & 0.96 \\
1000 & -40.75 & 0.56&   2.33 &  0.92 & 0.12 & 0.91 \\
\end{tabular}
\end{ruledtabular}
\label{tab.bks1f}  
\end{table}
\begin{table}     
\caption{ESC04a $^3S_1-^3D_1(\Xi N\rightarrow \Xi N, I=0)$ 
         BKS-phase parameters in [degrees] as a function of the laboratory 
         momentum $p_\Xi$ in [MeV]}
\begin{ruledtabular}
\begin{tabular}{r|rrr|rrr}
$p_{\Xi}$  & \multicolumn{1}{c}{$\delta(^3S_1)$}&\multicolumn{1}{c}{$\epsilon_1$} 
           & \multicolumn{1}{c|}{$\delta(^3D_1)$} 
             & \multicolumn{1}{c}{$\eta_{11}$}&\multicolumn{1}{c}{$\eta_{12}$} 
             & \multicolumn{1}{c}{$\eta_{22}$}\\
\colrule   
 10  &   2.02 & 0.00 &   0.00 &  1.000 & 0.000 & 1.000 \\
 50  &   9.75 &  0.00&   0.00 &  1.000 & 0.000 & 1.000 \\
 100 &  17.71 &  0.00&   0.00 &  1.000 & 0.004 & 1.000 \\
 200 &  25.97 &  0.00&   0.06 &  1.000 & 0.011 & 1.000 \\
 300 &  26.67 &  0.00&   0.19 &  1.000 & 0.010 & 1.000 \\
 400 &  23.53 &  0.00&   0.32 &  1.000 & 0.001 & 1.000 \\
 500 &  18.63 & 0.00 &   0.31 &  1.000 &--0.010 & 1.000 \\
 550 &  15.85 & 0.00 &   0.21 &  1.000 &--0.015 & 1.000 \\
 600 &  12.95 & 0.00 &   0.04 &  1.000 &--0.020 & 1.000 \\
 700 &   6.97 & 0.00 & --0.59 &  1.000 &--0.027 & 1.000 \\
 800 &   0.89 & 0.00 & --1.61 &  0.999 &--0.032 & 0.999 \\
 900 & --5.16 & 0.00 & --3.01 &  0.999 &--0.034 & 0.999 \\
1000 & --11.11& 0.00 & --4.71 &  0.999 &--0.033 & 0.999 \\
\end{tabular}
\end{ruledtabular}
\label{tab.bks1g}  
\end{table}
\begin{table}     
\caption{ESC04d $^3S_1-^3D_1(\Xi N\rightarrow \Xi N, I=0)$ 
         BKS-phase parameters in [degrees] as a function of the laboratory 
         momentum $p_\Xi$ in [MeV]}
\begin{ruledtabular}
\begin{tabular}{r|rrr|rrr}
$p_{\Xi}$  & \multicolumn{1}{c}{$\delta(^3S_1)$}&\multicolumn{1}{c}{$\epsilon_1$} 
           & \multicolumn{1}{c|}{$\delta(^3D_1)$} 
           & \multicolumn{1}{c}{$\eta_{11}$}&\multicolumn{1}{c}{$\eta_{12}$} 
           & \multicolumn{1}{c}{$\eta_{22}$}\\
\colrule   
 10  & 110.06 & 0.00 &  0.00 &  1.000& 0.000 & 1.000\\
 50  &  88.14 & 0.00 &  0.00 &  1.000& 0.000 & 1.000\\
 100 &  79.61 & 0.00 &  0.00 &  1.000& 0.002 & 1.000\\
 200 &  66.14 & 0.00 &  0.02 &  1.000& 0.006 & 1.000\\
 300 &  53.94 & 0.00 &  0.14 &  1.000& 0.015 & 1.000\\
 400 &  42.57 & 0.00 &  0.42 &  1.000& 0.028 & 1.000\\
 500 &  31.96 & 0.00 &  0.87 &  0.999& 0.046 & 0.999\\
 550 &  26.91 & 0.00 &  1.13 &  0.998& 0.057 & 0.998\\
 600 &  22.04 & 0.00 &  1.38 &  0.998& 0.068 & 0.998\\
 700 &  12.73 & 0.00 &  1.78 &  0.996& 0.092 & 0.996\\
 800 &   3.96 & 0.00 &  1.92 &  0.993& 0.117 & 0.993\\
 900 & --4.36 & 0.00 &  1.69 &  0.990& 0.140 & 0.990\\
1000 &--12.27 & 0.00 &  1.07 &  0.987& 0.161 & 0.987\\
\end{tabular}
\end{ruledtabular}
\label{tab.bks1h}  
\end{table}

 \end{widetext}






\begin{thebibliography}{99}
\bibitem{Rij05} Th.A.\ Rijken, {\it Extended-soft-core Baryon-Baryon Model.
                I, Nucleon-Nucleon Interactions}, 
                Phys. Rev. {\bf C73}, 044007 (2006) [arXiv:nucl-th/0603041]
\bibitem{RY05}  Th.A.\ Rijken and Y.\ Yamamoto, 
                {\it Extended-soft-core Baryon-Baryon Model. 
                II, Hyperon-Nucleon Interactions},
                Phys. Rev. {\bf C73}, 044008 (2006) [arXiv:nucl-th/0603042]
\bibitem{SR99}  V.G.J.\ Stoks and Th.A.\ Rijken, 
                Phys.\ Rev.\ C {\bf 59}, 3009 (1999).
\bibitem{Tak01} H.\ Takahashi, et al., Phys.\ Rev.\ Lett.\ {\bf 87}, 
                212502 (2001).
\bibitem{Dan63} M.\ Danysz, et al., Nucl.\ Phys.\ {\bf 49} 121 (1963).
\bibitem{Pro66} D.\ J.\ Prowse, Phys.\ Rev.\ Lett.\ {\bf 17} 782 (1966).


\bibitem{RSY99} Th.A.\ Rijken, V.G.J.\ Stoks, and Y.\ Yamamoto,
         Phys.\ Rev.\ C {\bf 59}, 1 (1999).
\bibitem{MRS89} P.M.M.\ Maessen, Th.A.\ Rijken, and J.J.\ de Swart,
         Phys.\ Rev.\ C {\bf 40}, 2226 (1989).

\bibitem{Bry81} R.A.\ Bryan, Phys.\ Rev.\ C {\bf 24}, 2659 (1981); {\bf 30} 305 (1984). 
\bibitem{Kla83} S.\ Klarsfeld, Phys.\ Lett.\ {\bf 126b}, 148 (1983).                    
\bibitem{Spr85} D.W.L.\ Sprung, Phys.\ Rev.\ C {\bf 32}, 699 (1985).                  
\bibitem{Kab87} A.\ Kabir and M.W.\ Kermode, J.Phys.G:Nucl.Phys. {\bf 13} (1987) 501.          

\bibitem{Nag73} M.M.\ Nagels, T.A.\ Rijken, and J.J.\ de Swart,
         Ann.\ Phys.\ (N.Y.) {\bf 79}, 338 (1973).

\bibitem{Mae95} P.M.M.\ Maessen, private communication.             


\bibitem{Swa63} J.J.\ de Swart,
         Rev.\ Mod.\ Phys.\ {\bf 35}, 916 (1963);
         {\bf 37}, 326(E) (1965).
\bibitem{Con35} E.U.\ Condon and G.H.\ Shortley,
         {\it The Theory of Atomic Spectra}
         (Cambridge University Press, Cambridge, England, 1935).

\bibitem{Tho70} R.H.\ Thompson,
         Phys.\ Rev.\ D {\bf 1}, 110 (1970).
\bibitem{Rij91} Th.A.\ Rijken,
         Ann.\ Phys.\ (N.Y.) {\bf 208}, 253 (1991).
\bibitem{Rij92} Th.A.\ Rijken and V.G.J.\ Stoks,
         Phys.\ Rev.\ C {\bf 46}, 73 (1992);
         {\bf 46}, 102 (1992).
\bibitem{Rij96} Th.A.\ Rijken and V.G.J.\ Stoks,
         Phys.\ Rev.\ C {\bf 54}, 2851 (1996);
         {\bf 54}, 2869 (1996).



\bibitem{Swa71} J.J.\ de Swart, M.M.\ Nagels, T.A.\ Rijken, and
         P.A.\ Verhoeven,
         Springer Tracts Mod.\ Phys.\ {\bf 60}, 138 (1971).
 \bibitem{Swa62} J.J.\ de Swart and C.K.\ Iddings, Phys.\ Rev.\ {\bf 128}, 
          2810 (1962); {\it ibid}\ {\bf 130}, 319 (1963).

 \bibitem{Tan65} Y.C.\ Tang and B.C.\ Herndon,
          Phys.\ Rev.\ {\bf 138}, B637 (1965).
 \bibitem{Bod65} A.R.\ Bodmer and S.\ Ali,
          Phys.\ Rev.\ {\bf 138}, B644 (1965).
 \bibitem{Dal89} R.H.\ Dalitz, D.H.\ Davis, P.H.\ Fowler, A.\ Montwill,
          J.\ Pniewski, and J.A.\ Zakrewski,
          Proc.\ Roy.\ Soc.\ (London) {\bf A426}, 1 (1989).
 \bibitem{SYM57} H.P.\ Stapp, T.\ Ypsilantis, and N.\ Metropolis, 
          Phys.\ Rev.\ {\bf 105},302 (1957).




\bibitem{E885}
P.~Khaustov {\it et al.}, Phys. Rev. {\bf C 61}, 054603 (2000).

\bibitem{NRS77}  M.M.\ Nagels, T.A.\ Rijken, and J.J.\ deSwart, 
Phys.\ Rev.\ D 15 (1977) 2547.


\bibitem{Yam94}
Y.Yamamoto, T.Motoba, H.Himeno, K.Ikeda and S.Nagata,
Prog. Theor. Phys. Suppl. {\bf No.117} (1994), 361.

\bibitem{Lanskoy}
D.E. Lanskoy, private communication.

\bibitem{Skyrme3}
M.\ Beiner, H.\ Flocard, N.V.\ Giai, and P.\ Quentin,
Nucl. Phys. {\bf A238}, 29 (1975)

\bibitem{Yamada92}
T.\ Yamada and K.\ Ikeda,
Prog. Theor. Phys. {\bf 88}, 139 (1992)

\bibitem{Ishikawa}
S.\ Ishikawa, M.\ Tanifuji, Y.\ Iseri and Y.\ Yamamoto,
Phys. Rev. {\bf C72}, 027601 (2005).

\bibitem{Ehime01} 
M.\ Yamaguchi, K.\ Tominaga, Y.\ Yamamoto, and T.\ Ueda,
Prog. Theor. Phys. {\bf 105}, 627 (2001)



\end{thebibliography}
\end{document}